%% file: main.tex
\documentclass[pra,aps,twocolumn,showpacs,superscriptaddress,nofootinbib,floatfix]{revtex4-2}
\usepackage{graphicx,amsfonts,amssymb,amsmath,mathrsfs,bm,dsfont}
\usepackage{braket}
\usepackage{anyfontsize}
\usepackage{mathtools}
\usepackage{physics}
\usepackage{stmaryrd}
\usepackage{appendix}
\usepackage{minitoc}

\usepackage[a4paper, portrait,text={17cm,24cm},centering]{geometry}
\usepackage{multirow}
\usepackage{hhline}
\usepackage{hyperref}
\hypersetup{ 
    colorlinks=true,
    linkcolor=blue,
    breaklinks=true,
    citecolor=blue,
    filecolor=blue,
    urlcolor=blue}
\usepackage{url}

\newcommand{\thetab}{{\boldsymbol{\theta}}}
\newcommand{\zb}{{\boldsymbol{z}}}
\newcommand{\Graph}{{\mathcal{G}}}
\newcommand{\edges}{{\mathcal{E}}}
\newcommand{\vertices}{{\mathcal{V}}}

\newcommand{\Cb}{\boldsymbol{C}}

\date{\today}

\begin{document}

\title{Implementing transferable annealing protocols \\for combinatorial optimisation on neutral atom quantum processors:\\ a case study on smart-charging of electric vehicles}

\author{Lucas Leclerc}
\email{lucas.leclerc@pasqal.com}
\affiliation{Pasqal, 24 rue Emile Baudot, 91120 Palaiseau, France}
\author{Constantin Dalyac}
\affiliation{Pasqal, 24 rue Emile Baudot, 91120 Palaiseau, France}
\author{Pascale Bendotti}
\affiliation{EDF R\&D, 7, boulevard Gaspard Monge, 91120 Palaiseau, France}
\author{Rodolphe Griset}
\affiliation{EDF R\&D, 7, boulevard Gaspard Monge, 91120 Palaiseau, France}
\author{Joseph Mikael}
\affiliation{EDF R\&D, 7, boulevard Gaspard Monge, 91120 Palaiseau, France}
\author{Loic Henriet}
\affiliation{Pasqal, 24 rue Emile Baudot, 91120 Palaiseau, France}

\begin{abstract}
In the quantum optimisation paradigm, variational quantum algorithms (VQA) face challenges with hardware-specific and instance-dependent parameter tuning, which can lead to computational inefficiencies. The promising potential of parameter transferability across problem instances with similar local structures has been demonstrated in the context of the Quantum Approximate Optimisation Algorithm. In this paper, we build on these advancements by extending the concept to annealing-based protocols, employing Bayesian optimisation to design robust quasi-adiabatic schedules. Our study reveals that, for Maximum Independent Set problems on graph families with shared geometries, optimal parameters naturally concentrate, enabling efficient transferability between similar instances and from smaller to larger ones. Experimental results on the Orion Alpha platform validate the effectiveness of our approach, scaling to problems with up to $100$ qubits. We apply this method to address a smart-charging optimisation problem on a real dataset. These findings highlight a scalable, resource-efficient path for hybrid optimisation strategies applicable in real-world scenarios. 
\end{abstract}

\maketitle 

\section*{Introduction}

Quantum Processing Units (QPUs) have garnered significant attention and investment for their potential to provide transformative speed-ups on complex problems~\cite{shor1994algorithms,grover1997quantum}. While current hardware capabilities are still maturing, this has spurred focused research on small-scale optimisation problems, laying a strong foundation for future scalability. Variational approaches have emerged as a powerful framework for leveraging quantum systems, guiding their dynamics toward final states that encode problem solutions~\cite{das2008colloquium}. The Quantum Approximate Optimisation Algorithm (QAOA)~\cite{QAOA,Blekos_2024,Zhou20} relies on a Trotterised version of an adiabatic evolution and has been extensively studied, especially in the quantum circuits' framework. More recently,  Variational Quantum Annealing Algorithms (VQAA) have been introduced as a way to design schedules capable of dynamically adapting to the gap structure of the Hamiltonian under consideration~\cite{zeng2016schedule,farhi2002quantum,perdomo2011study, crosson2014different}. These approaches include innovative techniques such as varying annealing speeds to optimise parameter regions or utilizing diabatic paths to circumvent the constraints of adiabatic evolution. Neutral atom platforms, with their unique advantages, are increasingly recognised as a robust resource for addressing graph-based optimisation problems, demonstrating exciting potential for practical quantum applications~\cite{henriet2020quantum, dalyac2024graph}.

 A noticeable challenge in deploying variational quantum optimisation algorithms effectively lies in the fine-tuning of algorithm parameters. Such tuning often needs to be tailored to specific problem instances and hardware characteristics, which introduces substantial overhead. This dependency hampers the general applicability and scalability of quantum optimisation approaches, as parameters that work well on one problem instance or hardware configuration may perform poorly on another. The issue becomes even more pronounced as quantum processors evolve, with unique device-specific limitations and error profiles that demand individualised parameter adjustments.

To answer this challenge, recent studies have explored the transferability of optimal parameters \cite{Pan2010-go,jiang2022transferabilitydeeplearningsurvey} across instances in the VQA paradigm, particularly in the QAOA framework\,\cite{montanezbarrera2024transferlearningoptimalqaoa}. A set of parameters $\thetab^*_C$, obtained as the optimal solution for a specific problem $C$, can be deemed transferable to another scenario $C^\prime$ if it retains its effectiveness, i.e. $C^\prime(\thetab^*_C)\approx \min_\thetab C^\prime(\thetab)$. $C$ and $C^\prime$ usually relates to each other by some shared structure, properties, or constraints. The smoothness, curvature, and presence of symmetries \cite{Lyngfelt2025-dg} in the optimisation landscape can also affect transferability as optimal sets from various instances concentrate in similar regions of the parameter landscape. The concentration of optimal QAOA parameters when solving graph-related problems, and the resulting successful transferability of these parameters across different problem instances, can usually be understood and predicted based on the local properties and subcomponents of the graphs \cite{galda2021transferabilityoptimalqaoaparameters}. An increased adaptability of QAOA would potentially minimise the need for exhaustive parameter re-optimisation across varying problem instances. 
In this work, we extend the investigation of parameter transferability to the VQAA framework, building upon Bayesian optimisation techniques\,\cite{leclerc:tel-04745992,PhysRevResearch.6.023063} to construct transferable control protocols. By analysing how parameter choices in quantum annealing can be adapted or transferred across different graph instances and hardware configurations, we aim to establish a more robust framework for quantum optimisation on graph families sharing similar structures. We also describe an implementation strategy that incorporates this transferability into the annealing process, with the goal of streamlining parameter optimisation and enhancing performance consistency across diverse instances.

\begin{figure*}[!]
    \centering
    \includegraphics[width=0.9\textwidth]{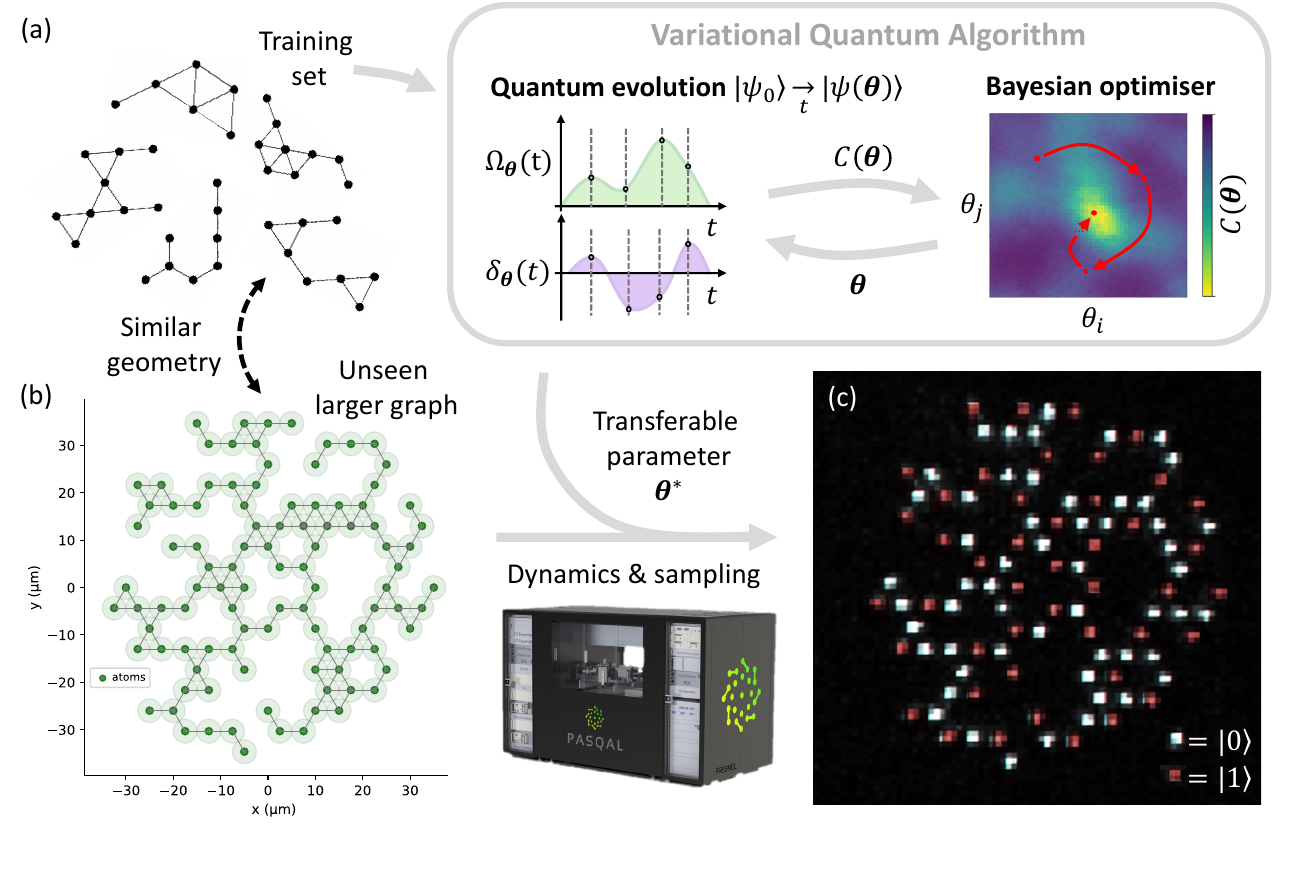}
    \caption{\textbf{Procedure to devise a transferable MIS preparation protocol.} (a) A training set of graphs, sharing the same geometry, are fed to a variational quantum algorithm tasked to find an optimal set of parameter $\thetab^*$  minimising $C(\thetab)$ for all graphs of the dataset. The variational loop consists in iteratively applying time-dependent control schedules $\Omega_\thetab(t),\delta_\thetab(t)$ to produce $\ket{\psi(\thetab)}$ and feed the measured $C(\thetab)$ to a classical Bayesian optimiser. The latter output new guesses $\thetab$ to try until convergence. (b) The optimised parameter set $\thetab^*$ can be applied to larger unseen graphs sharing the same geometry as graphs of the training set. A graph can be embedded in an atomic register (dark green dots) using the Rydberg blockade effect (lighter green) which enables to reproduce its connectivity and drives the quantum dynamics happening in a neutral atom device like Orion-alpha. (c) Experimental fluorescence picture obtained when sampling a MIS configuration of a $100$-qubit system. Atoms are either recaptured and measured in $\ket{0}$ (white dots) or expelled and measured in $\ket{1}$ (red dots), showcasing one MIS instance.}
    \label{fig:fig_1}
\end{figure*}

This article is structured as follows. In Sec.\,\ref{sec:ingredients}, we describe our specific implementation of a VQA to solve Maximum Independent Set (MIS) problems on an Orion class device, a Rydberg atom quantum processor from Pasqal. In Sec.\,\ref{sec:transferability},  we study the concentration of optimal parameters for specific families of graphs and introduce a method to devise transferable control protocols applicable to unseen instances of the same family. In Sec.~\ref{sec:applications}, we demonstrate how, within the context of a constrained budget of QPU access, the previously described method can be effectively applied to experimentally tackle the resolution of an industrial smart-charging problem. Finally, in Sec.~\ref{sec:scaling}, we experimentally investigate how the efficiency of such a protocol, as well as the quality of the solutions it produces, scale with the size of the graphs to which it is applied. 

\section{Ingredients and implementation of an analog VQA}
\label{sec:ingredients}
Variational quantum algorithms (VQAs) are hybrid classical-quantum procedures that iteratively minimise a cost function $C$. As illustrated in Fig.,\ref{fig:fig_1}(a), a VQA alternates between a classical optimisation routine, which adjusts a parameter $\thetab$ within a space $\Theta$, and a quantum processor, used to evaluate $C(\thetab)$. Upon reaching convergence or exhausting the call budget, the algorithm outputs the optimised parameter $\thetab^*$. In the following, we describe the four key components necessary for implementing the VQAs utilised in this study.

In the context of this paper, we focus on solving a particular graph optimisation problem, i.e. finding Maximum Independent Sets (MISs) of a unit disk (UD) graph. The UD-MIS problem involves finding the largest set of vertices in a graph such that no two are connected, while UD graphs represent spatial graphs where vertices are connected if they are within a fixed distance of each other, which can natively be embedded on a neutral atom quantum processor. 

\subsection{Quantum dynamics on a Rydberg processor}
\label{ssec:dynamics}
The quantum processor lies at the core of a VQA, performing the computationally intensive task of preparing potential solution states \(\ket{\psi(\thetab)}\), from which samples are drawn and the cost function computed. Here, we use a neutral-atom based QPU made of single $^{87}$Rb atoms trapped in arrays of optical tweezers~\cite{barredo2018synthetic,Nogrette14,browaeys2020many,henriet2020quantum,Morgado2021}. Qubits are encoded into the ground state $\ket{0}=\ket{5S_{1/2},F=2,m_F=2}$ and a Rydberg state $\ket{1}=\ket{60S_{1/2}, m_J=1/2}$ of the atoms.
The dynamics of a collection of $N$ qubits at positions $\boldsymbol{r}$ are governed by the following Hamiltonian:
\begin{equation}
    \hat H[\thetab](t;\boldsymbol{r})=\frac{\hbar\Omega_{\thetab}(t)}{2}\sum_{i=1}^N\hat\sigma^x_i-\hbar\delta_{\thetab}(t)\sum_{i=1}^N \hat{n}_i +\hat U(\boldsymbol{r}) 
    \label{eq:ham}
\end{equation}
where $\hat{\sigma}_i^\alpha$
are Pauli matrices and $\hat n_i=\ket{1}_i\bra{1}$. The two parameterisable time-dependent control fields are the Rabi frequency  $\Omega_\thetab(t)$ and the detuning $\delta_\thetab(t)$. $\hat U(\boldsymbol{r})=\sum_{i<j}C_6|r_i-r_j|^{-6}\hat{n}_i\hat{n}_j$ is the position-dependent interaction operator, with $C_6/h=138~$GHz$\cdot \mu {\rm m}^6$ for the Rydberg state considered \cite{Beguin_2013,ibali2017}. 
Evolving the system, initially in $\ket{\psi_0}$, according to $\hat H[\thetab]$ produces $\ket{\psi(\thetab)}=\sum_{\zb\in\mathbb{B}^N} a_{\boldsymbol{z}}(\thetab)\ket{\zb}$, with $a_\zb$, complex amplitudes of each bitstring $\zb$. Measuring the system provides one bitstring $\zb$ with probability $|a_\zb(\thetab)|^2$ and the distribution can be reconstructed by accumulating shots, i.e. repeating several times the preparation and measurement procedure.  

\subsection{Cost evaluation for UD-MIS}

Given a graph $\Graph=(\vertices,\edges)$, solving the MIS problem consists in finding
the largest subset $S\subset\vertices$ of vertices which do not share an edge in $\edges$. Such constraints can be encoded in the following cost Hamiltonian \cite{pichler_MIS}:
\begin{equation}
    \hat C_\Graph=-\sum_{i\in\vertices}\hat n_i+c\hat H_\Graph,
\end{equation}
with $\hat{H}_\Graph=\sum_{(i,j)\in\edges}\hat{n}_i\hat{n}_j$ encompassing the connectivity of $\Graph$ and $c\gg1$. 
Exploiting the representation of $\hat H_\Graph$ within the interaction Hamiltonian $\hat U$ of Rydberg atoms has spurred the development of various proposals for solving MIS problems~\cite{pichler_MIS,dalyac2024graph}. Most rely on preparing low energy states of the system described by Eq.\,\ref{eq:ham} as $\hat C_\Graph$ can be mimicked from $\hat H$ by arranging a geometric configuration $\boldsymbol{r}_\Graph$ of atoms such that $\hat U(\boldsymbol{r}_\Graph)\propto\hat H_\Graph$ (see App.\,\ref{app:mapping} for the detailed mapping). An example of embedding a $100-$node graph in an atomic register is displayed in Fig.\,\ref{fig:fig_1}(b). Solving a MIS problem ultimately boils down to identifying a control protocol $\Omega_\thetab(t),\delta_\thetab(t)$ efficiently preparing low energy states of $\hat C_\Graph$, i.e. the state minimising its energy. Once the system has been driven to such $\ket{\psi(\thetab)}$, evaluating the cost observable $\hat C$ returns $C(\thetab)=\langle\psi(\thetab)|\hat C|\psi(\thetab)\rangle=\sum_{\zb\in\mathbb{B}^N} |a_{\zb}(\thetab)|^2\langle\zb|\hat C|\zb\rangle$. Thus, $C(\thetab)$ can directly be reconstructed from the probability distributions $|a_\zb(\thetab)|^2$ measured on the quantum processor. Fig.\,\ref{fig:fig_1}(c) showcases the experimental picture of the atomic system embedding the graph of (b) when measured in a MIS configuration. 

In the following, we will use two cost operators, leading to two distinct cost functions favouring MIS configurations. A straightforward procedure consists in only looking at the proportion of MISs in the measured distribution (and flip it to lower the cost), i.e.~$1-$P$(\text{MIS})$. The required MIS size $S_\Graph$ must be determined classically, either through an exhaustive search feasible for $|\vertices|\lesssim 50$ or via reliable classical optimisation methods (detailed in \ref{ssec:sampling}) for larger system sizes. For a bitstring $\zb$ to be considered a MIS, it must satisfy the condition $\langle\zb|\hat C_\Graph|\zb\rangle=-S_\Graph$. Alternatively, evaluating $\hat C_\Graph$ on the whole distribution gives the approximation ratio, a more nuanced measure that accounts for the contributions of large IS in reducing the total cost, overcoming the binary projection issue of the first choice. We define the normalised approximation ratio $R_\Graph(\thetab)=\langle\psi(\thetab)|\hat C_\Graph|\psi(\thetab)\rangle/S_\Graph$. Inside the VQA loop, the evaluated cost $C(\thetab)$ is then passed to the classical optimisation routine to guide the search for improved controls at the next iteration.

\subsection{Bayesian optimisation for parameter space navigation}
Gradient-based optimisation methods can handle parametrisations with hundreds of variables. However, the presence of barren plateaus \cite{mcclean2018barren,Holmes_2022,Larocca_2022}, local minima and the high cost of estimating unknown gradients for noisy expensive-to-evaluate black-box functions \cite{Schuld_2019,Izmaylov_2021} limit their success in VQAs. By using ansätze with fewer parameters (typically fewer than $20$), gradient-free methods like Bayesian optimisation (BO) \cite{brochu2010tutorial} become viable. This approach utilises a statistical model to approximate the cost function $C$ while quantifying uncertainty, and an acquisition function to guide the search. After an initial training phase, subsequent iterations balance exploration and exploitation, helpful for getting out of local minima, updating the model with each new evaluation through Bayesian inference as demonstrated in App.\,\ref{app:bayesian-optimisation}. This method is particularly advantageous when the number of iterations needs to be kept low and when dealing with noisy evaluations, such as those arising from low sampling of final states on a quantum computer \cite{PRXQuantum.1.020322}. In addition, with the right parametrisation, it can constitute an effective pathfinder when designing quantum annealing schedules, notably for the $p-$spin model \cite{Fin_gar_2024}. 

\subsection{Control parametrisation: QAOA \& VQAA}

A problem-informed parametrisation of the controls plays a crucial role in creating a more navigable optimisation landscape. A standard approach is using a schedule of external controls to guide the system from an easy-to-prepare ground state of an initial Hamiltonian to the desired, more complex, ground state of a final Hamiltonian, say $\hat C_\Graph$ in our case. This approach relies on the adiabatic theorem (see App.\,\ref{app:adiabatic}), which in essence ensures that if a system evolves slowly compared to its characteristic energy scale, it follows the instantaneous eigenstate of the Hamiltonian. 

In the VQAA framework~\cite{Farhi_2001}, the parametrisation needs to create continuous quasi-adiabatic schedules. $\thetab$ encompasses the values of the two control fields at fixed points in time, i.e. $\{\Omega_{\thetab}(t_i),\delta_{\thetab}(t_i)\}_{i=1\cdots m}$, as well as the total evolution duration $T$. In this implementation, the $t_i=T(i-1)/(m-1)$ are fixed, but more generally they can be focused in regions where one needs more flexibility in the evolution, or they can even be set as optimisation parameters. The parameter space can be shrunk to a ($2m+3$)-dimensional one by fixing $\Omega_{\thetab}(0)=\Omega_{\thetab}(T)=0$ to account for hardware constraints. The control fields are finally derived using interpolation by monotonic cubic splines, which ensures that intermediate values will not cross the control bounds, as showcased in an example in Fig.\,\ref{fig:optim-pulse}. This parametrisation has the advantage of defining schedules of arbitrary duration with relatively few parameters, on the contrary to the QAOA one.

In the QAOA framework, the parametrisation relies on the depth $p$ with the procedure alternating $p$ times between two distinct operations: a mixing stage and a problem-specific evolution. $\thetab$ effectively encompasses the duration of each of the control layers (see App.\,\ref{app:qaoa} for more details). Each layer thus includes the period of free evolution with constant detuning $\delta$ lasting $t_{\rm cost}$ followed by a resonant pulse with constant amplitude $\Omega$ lasting $t_{\rm mix}$. The sequence is usually initialised with a $\pi/2$ resonant pulse which, without interactions, would prepare the system in $\ket{+}^{\otimes N}$ but here, only spreads the state as much as possible. As increased performance usually requires higher depth, the parameter space can grow significantly, stalling the usefulness of BO or requiring clever initialisation strategies, such as sequential optimal parameters transfers from depth $p$ to $p+1$ \cite{Pichler2018}. 

The controls can be built using \texttt{Parameterized Sequence} objects in the Pulser library \cite{silverio2022pulser} and are  sent to either numerical emulators or remotely to the Pasqal device through a cloud platform, closing the iterative loop started in \ref{ssec:dynamics}.

\section{Transferability of optimal parameters in variational quantum algorithms}
\label{sec:transferability}

Variational algorithms face significant challenges due to their computational expense and sensitivity to hardware conditions, motivating strategies to enhance efficiency and robustness. In this section, we explore the transferability of variational parameters for families of graphs generated on different geometries.

\subsection{Looping cost and the need for transferability}

VQAs are computationally costly to run due to their iterative feedback loops. Each additional iteration incurs substantial resource usage, particularly on current noisy quantum hardware where accuracy requires large numbers of shots to estimate observables, and thus $C(\thetab)$, reliably. Respecting a budget of $n_{\rm iter} \times n_{\rm shots}$ QPU calls imposes trade-offs between reducing the number of iterations or lowering measurement precision, both of which ultimately limit performance. Thus, Bayesian optimisation approaches which aim to minimise $n_{\rm iter}$ even with high shot noise (low $n_{\rm shots}$) are critical. Furthermore, the optimal parameters $\boldsymbol{\theta}^*$ found by VQAs can strongly depend on the specific quantum hardware configuration, necessitating full re-optimisation upon recalibration (App.~\ref{app:miscalibration}). This motivates developing parametrisation robust to hardware noise and control fluctuations, such as the VQAA demonstrated in App.~\ref{app:noise-VQAA}.

For the MIS problem on a dataset of $M$ graphs (Fig.~\ref{fig:fig_1}(a)), naively running VQAs independently for each graph requires a shot budget of $M \times n_{\rm iter} \times n_{\rm shots}$. Parameter transferability, which reuses $\boldsymbol{\theta}^*_{\mathcal{G}}$ optimised for graph $\mathcal{G}$ as a warm start or pre-optimised protocol for $\mathcal{G}'$, offers resource savings. Its success depends on the concentration of optimal parameters across graph families, which depends significantly on the similarity between graphs in terms of size, density, and local connectivity \cite{montanezbarrera2024transferlearningoptimalqaoa}.

This work (Fig.~\ref{fig:fig_1}) proposes a protocol to solve MIS for similar graph families by training on smaller instances. This avoids or efficiently warm-starts optimisation for larger graphs, such as the $100$-node example (Fig.~\ref{fig:fig_1}(b)). Training can be performed classically using emulated VQAs, focusing quantum resources on larger graphs to improve sampling efficiency. In the following, we elaborate on this idea by studying for the two parametrisations mentioned above the concentration of optimal parameters for several families of graphs.

\subsection{Cost landscapes for various geometries and parametrisations}
\label{ssec:geometries}
\begin{figure*}[!]
    \centering
    \includegraphics[width=\textwidth]{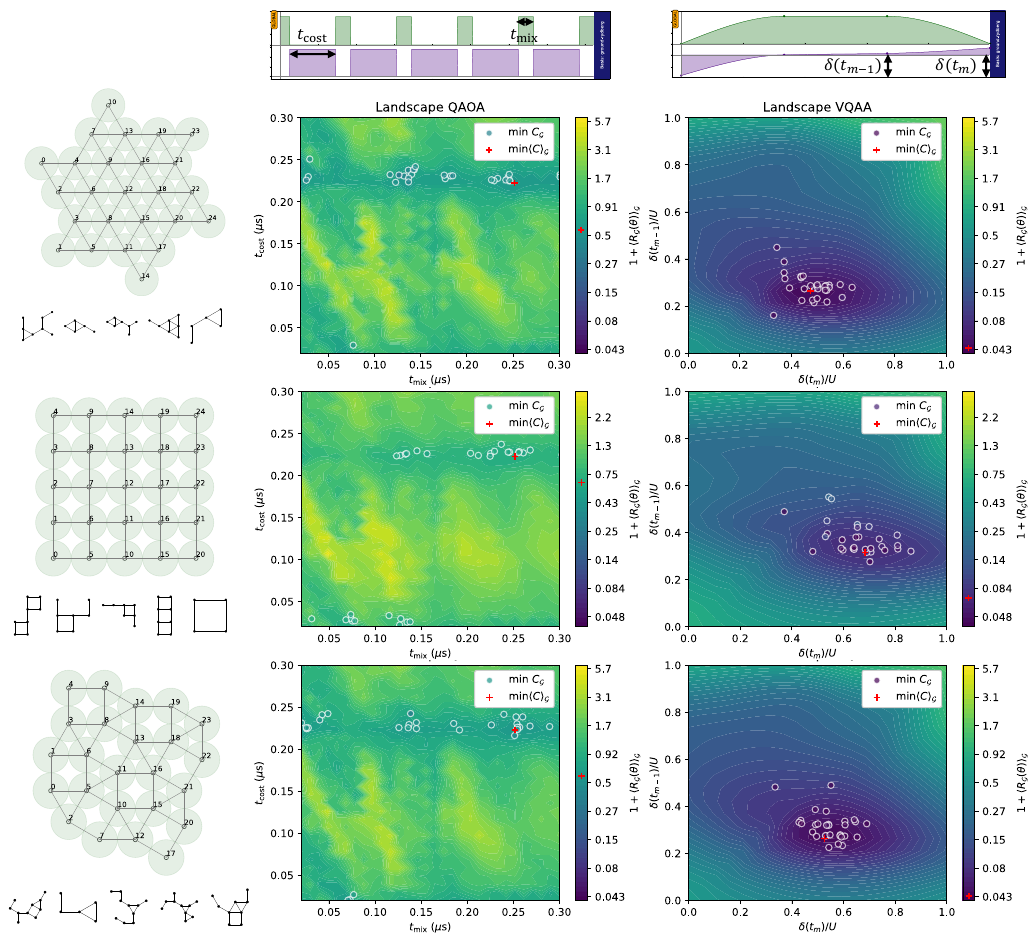}
    \caption{\textbf{Study of generalisability of VQAs over graphs generated from regular lattices.} Datasets of graphs generated from either triangular, square or Shastry-Sutherland lattices (layouts plotted on the left) are subjected to QAOA-like or VQAA-like protocols parametrised by two parameters (Pulser sequences plotted above). For the QAOA-like protocol, a depth-5 sequence is parameterized by the times $t_{\rm cost}$ and $t_{\rm mix}$ at depth $1$, with the steps at higher depths fixed to these same values. For the VQAA protocol, only the last two detuning values of a standard VQAA protocol are parameterised. The corresponding colour maps are obtained by averaging over the dataset the normalised approximation ratio $R_\Graph(\thetab)$ landscapes for each graph across the parameter space. The minimum found for the averaged landscape is plotted (red cross) as well as the minimum found for each graph over this space (dots). Regions with lower cost are displayed in darker colours.}
    \label{fig:landscape-geo}
\end{figure*}

\begin{figure*}[!]
    \centering\includegraphics[width=1.01\textwidth]{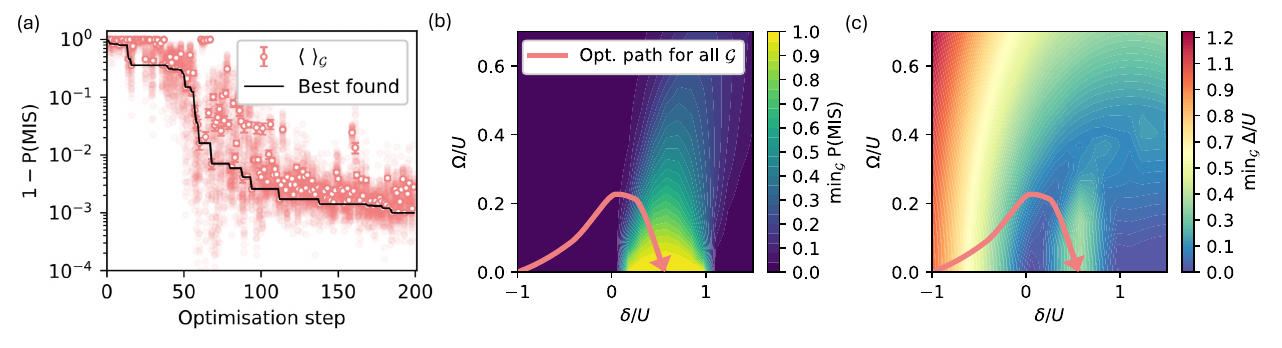}
    \caption{\textbf{Optimisation results when devising a unique VQAA protocol trained on a dataset of triangular graphs.} (a) Convergence of the Bayesian optimiser when searching for regions of parameters minimising $1-\langle$P(MIS)$\rangle_\Graph$ (dots) a VQAA parametrisation (light red). The cost obtained for each graph of the dataset at each iteration is plotted in transparent. The historical lowest cost found is also kept track of (black) (b) Phase diagram of the minimum MIS probability calculated across the training set. The map is obtained by computing the ground state of the Rydberg Hamiltonian of Eq.\,\ref{eq:ham} with constant parameters $\Omega/U$ and $\delta/U$  for each graph, measuring the associated P(MIS) and plotting the minimum across the $50$ graphs. An example of a P(MIS) landscape for a single graph is given in Fig.\,\ref{fig:optim-unique}(b). The path taken by the optimised protocol (light red) is constrained by hardware-realistic bounds, given in the text, but succeeds in locating the reduced MIS phase (yellow). (c) Similar as (b) for the minimum gap $\Delta$ across the $50$ graphs of the dataset. The optimised protocol avoids regions of vanishing gaps (blue). An example of a gap landscape for a single graph is given in Fig.\,\ref{fig:optim-unique}(c).}
    \label{fig:optim-general}
\end{figure*}

We focus on studying transferability for the MIS problem on families of similarly structured UD graphs. UD graphs are inherently local in structure 

A way of building a family of graphs with similar structure consists in generating instances by randomly sampling a regular lattice, using random walks. Given a periodic trap layout, we can construct families of graphs by selecting subsets of adjacent traps, either including or excluding cycles. This method ensures that the resulting graphs within a family share a similar local structure, as they inherit fundamental properties from the underlying lattice. More precisely, the maximum degree of any graph within such a family is constrained by the degree of the lattice it is derived from—for example, a maximum degree of 6 for triangular lattices and 4 for square lattices.
Additionally, the possible cycle lengths within these graphs are also dictated by the structure of the parent lattice. In a triangular lattice, the smallest permissible cycle lengths include 3, 6, and 8, while in a square lattice, they include 4, 8, and 10. This restriction arises because cycles must be composed of edges that align with the fundamental symmetries of the lattice, limiting the ways in which they can form. Consequently, graphs derived from the same lattice exhibit not only similar local connectivity patterns but also comparable combinatorial properties, reinforcing their classification into the same family. Examples of such families are given in the first column of Fig.\,\ref{fig:landscape-geo} using triangular geometry, square geometry or a combination of the two, taking the form of a periodic Shastry-Sutherland lattice, well-known in condensed-matter physics~\cite{Liu2014}. These geometries do not encompass the vast majority of UD graphs. Each layout has a minimum spacing of $5\mu$m leading to a maximum blockade strength of $U/2\pi\approx9$ MHz across the array. For each layout, we generate a family of $30$ graphs of size ranging from $N=5$ to $10$ and apply to each graph either a QAOA-like or a VQAA-like sequence. Each protocol, showcased on the first line of Fig.\,\ref{fig:landscape-geo}, has two parameters, and we obtain the corresponding landscapes of normalised approximation ratio by sweeping over their possible values within the bounds $\Theta=[0.02 \,\mu\rm{s},0.3\,\mu\rm{s}]^2$ for the QAOA-like protocol and $\Theta=[0,U]^2$ for the VQAA. The minimal duration of $0.02 \,\mu\rm{s}$ is set by the ramp time of the shaping device used to produce square pulses. Fig.\,\ref{fig:landscape-geo} depicts the cost landscapes, averaged over the graphs of each family, with additionally the location of the global minimum for each graph also indicated to highlight possible concentration effects and global efficiency discrepancy between the two approaches. 

In the QAOA-like case, the optimal parameters for each graph are strongly concentrated along the \( t_{\rm cost} \) direction, while they remain more spread along the \( t_{\rm mix} \) direction, suggesting less sensitivity to variations in the latter. Due to practical constraints in the implementation of the protocol, the local minima for each graph are not observed to approach $0$ in value. For each of the three geometries considered, the individual minima obtained for each graph only partially group together, invalidating the hypothesis of a single QAOA-like protocol enabling preparation of MIS over the graph families studied. Nevertheless, a limited concentration effect can still be observed, indicating that some degree of transferability of the optimal parameters arises between the studied graph families. Additionally, the landscapes between families remain similar, indicating a possible transfer of parameters between geometries.  

On the contrary, for the VQAA-like approach, the individual landscapes overlap to create a global minimum for the averaged case, closely surrounded by the clustered individual global minimum locations. For the triangular and Shastry-Sutherland layouts, the global minimum of the averaged landscape reaches $0.04$ whereas for the square case, it only approaches $0.06$. Two elements can contribute to explain this discrepancy : firstly, some graphs in this family may not be as similar as ones in the triangular family, which is supported by the existence of some individual minima located far from each other. Secondly, in the square case, the second next neighbours interactions are often only reduced by a factor $1/8$ compared to the next neighbour ones, making such graphs more prone to have imperfect embeddings. If $\hat H_\Graph$ is not well reproduced, a VQAA protocol will only yield a finite efficiency. This effect seems reduced when combining square and triangular geometries. In addition, we can also see that variations of the parameters will only poorly affect the averaged preparation, as the landscapes of VQAA remains quite flat around the global minima. Finally, it is interesting to notice that the optimal locations varies between the geometries with $\delta(t_m=T)$ increasing from triangular to square, passing by Shastry-Sutherland, reminiscing of condensed-matter phase diagrams of the antiferromagnetic regular lattices \cite{Lienhard2018}. This gives an interesting insight on the values around which the MIS phases of the various geometries are centred. The lattices studied in this section can not reproduce any graphs (for instance, graphs with cycles of length 5) and additional geometries, such as Kagome or King's lattice, could also be studied. Since our study focuses exclusively on regular 2D lattices with local connectivity—rather than highly connected graphs potentially extending toward complete graphs—the associated concentration regions tend to overlap. This suggests that while optimization paths need to be geometry-dependent, similarities between lattice structures may still allow for transferable protocols.  
Those observations motivate the choice of the VQAA approach to design in the next section a transferable control protocol.

\subsection{Devising a transferable protocol for a specific geometry}
\label{ssec:train}

The objective of this section is to derive a generalised schedule capable of preparing MIS configurations for all graphs within the triangular family, including those for which the protocol has not been specifically optimised on during the training. In the next section, the resulting protocol will then be applied directly to the testing set, comprising unseen, larger graphs

The optimisation is performed classically over a training set of $50$ small graphs, $10$ of each size ranging from $5$ to $9$, including instances with and without cycles. We detail in App.\,\ref{app:vQAA} the procedure and the strengths of this VQAA method for a unique graph. The cost $C(\thetab)$ is obtained summing the average $\langle~\rangle_\Graph$ of the MIS probability obtained for each graph and the associated standard deviation $\sigma_\Graph$ with a weight set here to $1$. Adding the standard deviation in the cost function helps the optimiser to favour protocols with uniform costs across the family over ones working very well for specific instances but less generalisable. Fig.\,\ref{fig:optim-general}(a) showcases the convergence over $n_{\rm iter}=25+175$ iterations (using $n_r=25$ iterations for initialisation of the BO algorithm, see App.\,\ref{app:bayesian-optimisation}) for a VQAA with $m=3$ and $T_{\rm max}=4~\mu$s. The VQAA locates an interesting region with cost below $0.5\%$ after $100$ iterations and ultimately converges towards $\langle$P(MIS)$\rangle_\Graph\approx99.73\%$, where all the graphs have a MIS probability above $98.5\%$. We can check that running separate optimisation routines for each graph gives very similar average convergence both in terms of speed and quality of preparation but with final costs more spread. The optimiser history can be boiled down to three main steps: before the $50^{th}$ iteration, it tries to work on all the graphs at the same time, while between $50$ and $125$, it allows itself to try pulses working very well, and even better than the final one, for only a few graphs while the cost of the others can be worse than before, as shown by the larger spread of individual scores on Fig.\,\ref{fig:optim-general}(a). Finally, after iteration $125$, it only refines the averaged cost by a few tens of $\%$. 

The optimised path given in Fig.\,\ref{fig:optim-general}(b) starts deep in the IS phase and ends in the middle of the MIS phase, where the MIS probability of the ground state goes to $1$ for all graphs. We have constrained the optimiser to use hardware realistic bounds ($\Omega_{\rm max}/2\pi=2~$MHz and $|\delta_{\rm max}|/2\pi\approx10~$MHz) explaining the smaller path taken compared to Fig.\,\ref{fig:optim-unique}(b). When taking the minimal value over the family of graphs, the MIS phase is narrower, i.e. $0.2\leq\delta/U\leq1$, aggregating the constraints from all the graphs of the dataset. Additionally, the optimiser identifies the optimal path while adhering to both hardware constraints and the adiabatic criteria, ensuring avoidance of regions with low energy gaps $\Delta$, for instance around $(\Omega,\delta)/U\approx (0,0)$, as shown in Fig.\,\ref{fig:optim-general}(c). Here, the gap $\Delta$ is defined between the MIS manifold of states and the first above excited state. In a 2D square lattice, the gap is minimal at the phase transition point and scales with the inverse of the system size $\Delta \sim 1 / \sqrt{N} $~\cite{sachdev1999quantum, schuler2016universal}. For the triangular lattice, a first-order quantum phase transition is anticipated from the paramagnetic phase to either the 1/3 or 2/3 filling crystals, leading to a minimal energy gap that decreases exponentially with N~\cite{laumann2015quantum, da2021phase}. It is therefore possible for this geometry to construct a unique optimised protocol, preparing MIS configurations with high efficiency for various graphs. In the following, we study its performance over a test set of unseen graphs, constructed from an industrial use case. 

\section{End-to-end pipeline for tackling smart charging tasks}
\label{sec:applications}

We will now apply the annealing protocol optimised above to address, with a neutral atom QPU, an industrial use case in the smart charging sector, provided by the French electric utility company EDF and numerically studied in~\cite{Dalyac_2021}. Specifically, we will use instances with constraints that can be naturally embedded on a triangular layout as a test set.

\begin{figure*}
    \centering
    \includegraphics[width=0.95\textwidth]{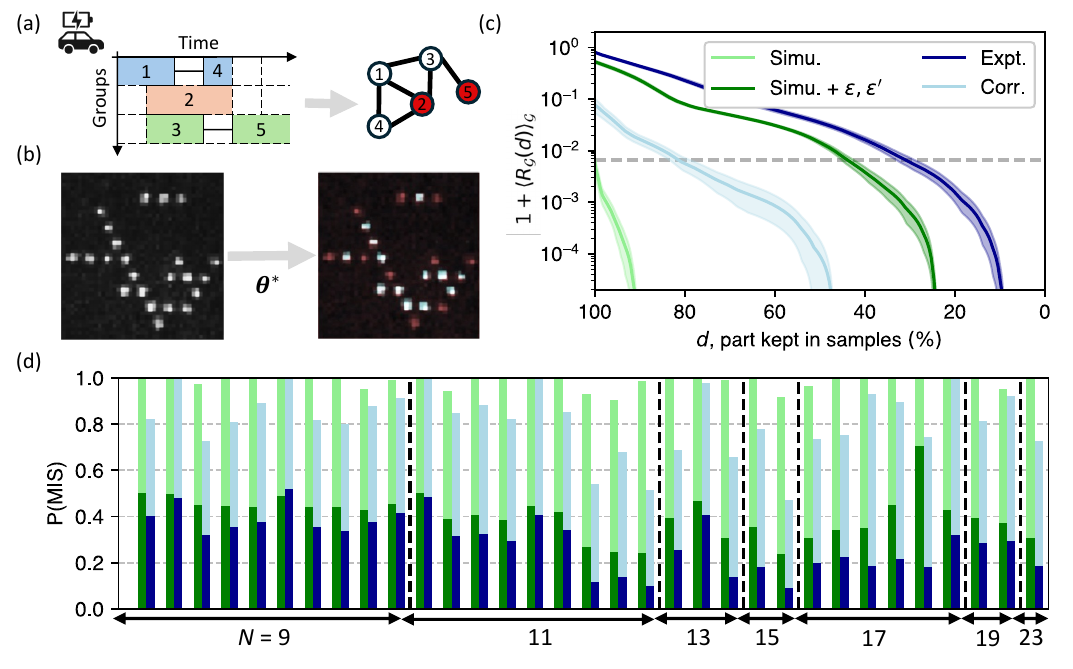}
    \caption{\textbf{Experimental results when applying the generalisable VQAA protocol to a smart charging use case.} (a) Example of a GISP with $3$ groups and $5$ loads, mappable to a MIS problem on a $5-$node graph. (b) Experimental fluorescence pictures of a $23-$atom register taken before and after the quantum dynamics. Atoms measured in Rydberg state (red) forms a MIS solution. (c) Evolution of the normalised truncated approximation ratio $\langle R_\Graph(d)\rangle_\Graph$, averaged over the dataset, when keeping only the $d\%$ best bitstrings of each distribution for the computation. Mean (solid line) and standard deviation (transparent filled areas) are plotted for (light green) noiseless emulation, (green) noisy emulation, experimentally obtained (blue) raw and (light blue) corrected data. (d) Histograms of the MIS probabilities obtained for each graph of the dataset.}
    \label{fig:exp-EDF}
\end{figure*}

\subsection{A use case in smart-charging}
\label{ssec:smart-charging}

Smart charging is an umbrella term encompassing all optimisation problems related to electric vehicle (EV) charging. The recent increase in the number of EVs presents both new challenges and opportunities for electricity management. Issues like charging task allocation, scheduling, and cost optimisation emerge due to the significant charging times of EVs and their unpredictable load on the electrical system~\cite{rafi2020comprehensive}. Vehicle batteries can serve as energy storage and power supply devices, a technique known as vehicle-to-grid, which significantly enhances the flexibility of the electric system and reduces the reliance on fossil fuels during high-peak demand~\cite{gschwendtner2021vehicle}.

Charging an EV takes a finite duration, with the starting time typically pre-scheduled by users. Each EV must be charged within a specific time interval and is assigned to a group, representing various company fleets or alternative time intervals for a single EV. With the advent of pre-booking systems in smart charging, users can select subsets of possible charging intervals, offering terminal operators some flexibility in assigning EVs to charging terminals. This concept of alternative intervals also arises in broader scheduling problems, such as energy management. For example, in nuclear outage planning~\cite{Griset22}, constraints include selecting one outage date from a set of time intervals for each production unit, effectively forming groups of alternatives.

Given a set of time intervals partitioned into groups, the objective is to serve the maximum number of EVs from distinct groups at a single terminal. This problem reduces to an Interval Scheduling Problem (ISP), where tasks correspond to intervals. Groups can be modelled in two ways: as subsets of tasks or as alternative intervals for a single task. Both approaches can be unified under the same problem formulation, with groups containing at most $K$ tasks. $K$ represents the maximum number of tasks allowed per group across all groups. A subset of tasks is compatible if no intervals overlap and no two intervals belong to the same group. This problem is formally defined as the Group Interval Scheduling Problem (GISP), which seeks the largest compatible subset of tasks. The GISP is NP-hard for $K\geq3$ and lacks a Polynomial-Time Approximation Scheme~\cite{Nakajima82}, but is polynomially solvable for $K\leq2$~\cite{Kiel92}. The GISP also reduces to finding the MIS on a Group Interval graph, as illustrated in Fig.\,\ref{fig:exp-EDF}(a) with a $K=2$ instance. In this formulation, each task corresponds to a vertex, with edges representing conflicts (incompatible tasks). Thus, solving the GISP for smart charging scenarios can be approached using the MIS-solving VQAA, provided the problem instances can be mapped to implementable graphs. Furthermore, we can apply the optimal parameter obtained in Sec.\,\ref{ssec:train}, provided the obtained graphs can be mapped to a triangular geometry.

\subsection{Experimental implementation and results}

The graphs used originate from a data set of $2250$ loads performed during May $2017$ on identical charging points of the Belib’s network of terminals located in Paris~\cite{Belib17}. The GISP instances can be derived by randomly sampling the loads, and we limit this work to specific instances that can be formulated as MIS on two-dimensional UD graphs.

We embed the $33$ graphs of sizes ranging from $N=9$ to $23$ into atomic registers with triangular geometry. Applying the transferable VQAA protocol obtained in the previous section returns, in noiseless emulation, $\langle$P(MIS)$\rangle_\Graph=98.3(4)\%$, a performance very close to the one of the training set. Running optimisation loops on some of those graphs would have required non-negligible classical resources in emulation, promoting the usefulness of the approach followed. Additionally, we experimentally apply the protocol on the hardware and imaging at the end of the dynamics, we can directly retrieve MIS configurations composed of the missing atoms between the two fluorescence pictures shown in Fig.\,\ref{fig:exp-EDF}(b). We acquire around $800$ shots for each graph and we overall get $\langle$P(MIS)$\rangle_\Graph=29(2)\%$. 

The general quality of the distributions, either emulated or experimentally obtained, can be summarised by looking at the evolution of the averaged truncated approximation ratio when discarding the worst bitstrings of the distributions.
Fig.\,\ref{fig:exp-EDF}(c) displays how the normalised approximation ratio approaches $1$ as bitstrings with high cost are discarded more and more. For instance, the averaged cost obtained in noiseless simulation (light green) is around $1\%$ but discarding the $10$ worst percent of the distributions makes it plummet to $0$, highlighting that the minimum MIS probability reached across the dataset is around $90\%$. The experimental curve (dark blue) drops at low $d$ emphasises that the worst preparation over the dataset still includes around $10\%$ of MISs in its distributions. We can also quantify how much the detection errors $\varepsilon=3\%$ and $\varepsilon^\prime=8\%$ (dark green) modify the costs by checking the proportion that needs to be discarded to retrieve the original cost (following the grey dashed line), here about $60\%$. The detection errors are not the only contributors to the discrepancies between emulated and experimental data, as correcting for those (light blue), following the procedure described in App.\,\ref{app:spam}, does not reproduce the emulated noiseless behaviour. Indeed, from Fig.\,\ref{fig:noise-analysis}(b), we expect from a $4~\mu$s annealing pulse a drop in MIS probability of at least $40\%$ due to decoherence, in addition to $20\%$ due to detection errors. Fig.\,\ref{fig:exp-EDF}(d) gives a more detailed picture of the experimental results with the smaller graphs ($N=9$) exhibiting between $35\%$ and $50\%$ of MIS configurations while few graphs at sizes $11$, $15$ and $19$ only have around $10\%$ of MISs. Correlating these significant performance drops with factors like the degeneracy of MIS in the graph remains challenging, as various experimental parameters may have also contributed to the observed outcomes. Although the devised protocol demonstrated resilience to control fluctuations, such as those caused by spatial inhomogeneities across the atomic array, potential synchronisation issues between the excitation lasers—stemming from the two-photon nature of the Rydberg transition employed—could have altered the time-dependent shape of the protocol.
Overall, for almost all graphs, except the ones with the largest MIS subspaces, such as the $23-$nodes instance ($32\%$ of the $264$ possible configurations found), all the MISs were found at least once. On average, the first MIS was typically identified within just a few shots—equivalent to seconds—demonstrating a performance improvement of several orders of magnitude compared to directly conducting closed-loop VQAs on each graph using the QPU, which usually require hundreds or thousands of shots at those sizes.

The size of the smart-charging instances was restricted to $N=23$ due to the challenges of encoding all constraints on a triangular layout as the instance size increases. Specifically, cliques larger than $3$ cannot be encoded within a triangular layout. In a GISP, there are two types of constraints: group constraints and overlapping interval constraints. Since some overlapping intervals inherently form triangles (i.e., cliques of size $3$), adding further group constraints becomes more and more impractical. As a result, we address only a small subset of practical industrial problems when scaling to larger sizes.
In the perspectives section, we will discuss potential approaches to encode more general constraints using ensembles of atoms. However, to evaluate the scalability and performance of our method on real hardware, we rely on an artificial dataset of graphs that fit on a triangular layout with sizes up to $N=100$ as presented in the next section.

\section{Scaling of MIS probability with graph size}
\label{sec:scaling}

The transferable protocol demonstrates comparable performance on a test set composed of graphs close in size to those of the training set. However, its reliance on the inherently adiabatic nature of the driving may hinder its efficiency when applied to larger instances, ultimately limiting scalability. In this section, we study how the proportions of solutions with different quality scales with the problem size, showing that good solutions can still be sampled at large sizes. 

\subsection{Sampling on the test set}
\label{ssec:sampling}
This second test set consists of $100$ triangular graphs, $10$ of each size ranging from $N = 10$ up to $N = 100$, obtained with the same random site sampling method as used in Sec.\,\ref{ssec:geometries}.  Emulating the dynamics over such large systems, especially for $N\geq25$, requires using tensor network methods described in \cite{bidzhiev2023cloud}. Another challenge in scaling arises from the classical computation of the MIS size $S_\Graph$ for each graph. We rely on two classical algorithms: an approximate algorithm using the subgraph exclusion (SE) method~\cite{Boppana1992} implemented in the \hyperlink{https://github.com/networkx/networkx?tab=readme-ov-file}{NetworkX} python package and the state-of-the-art \hyperlink{https://www.ibm.com/fr-fr/products/ilog-cplex-optimisation-studio/resources}{CPLEX} solver from IBM. CPLEX employs a variety of advanced optimisation techniques, including linear programming, branch-and-bound, cutting planes, and heuristics, to efficiently solve large-scale optimisation problems. We numerically apply the transferable protocol obtained in the previous section to all the graphs of the test set, and we check the size of the maximal IS found for each of them when sampling the final state $n_{\rm shots}=1000$ times.

For all graphs, we have $S_\Graph^{\rm SE}\leq S_\Graph^{\rm VQAA}\leq S_\Graph^{\rm CPLEX}$. The SE method proves to be unreliable as it can not even reach the MIS sizes found by an exhaustive search at graph size $\leq50$ and is always matched or exceeded by the general VQAA approach. When compared to the sizes obtained by CPLEX  (v$22.1.1$ executed in under $10~$ms on a dual AMD Rome 7742 system with 128 cores at 2.25GHz), the VQAA method correctly identifies the MIS size in $82\%$ of cases, deviating by just one node in the remaining instances, with more errors as the graph size increases. No variational optimisations were performed on these graphs; only the transferable annealing drive computed in Sec.\,\ref{ssec:train} was applied. MIS (or MIS-1) were found after only a few dozen of shots, thus with a time to solution around three orders of magnitude higher than with CPLEX. 

\begin{figure}[!]
    \centering
    \includegraphics[width=0.49\textwidth]{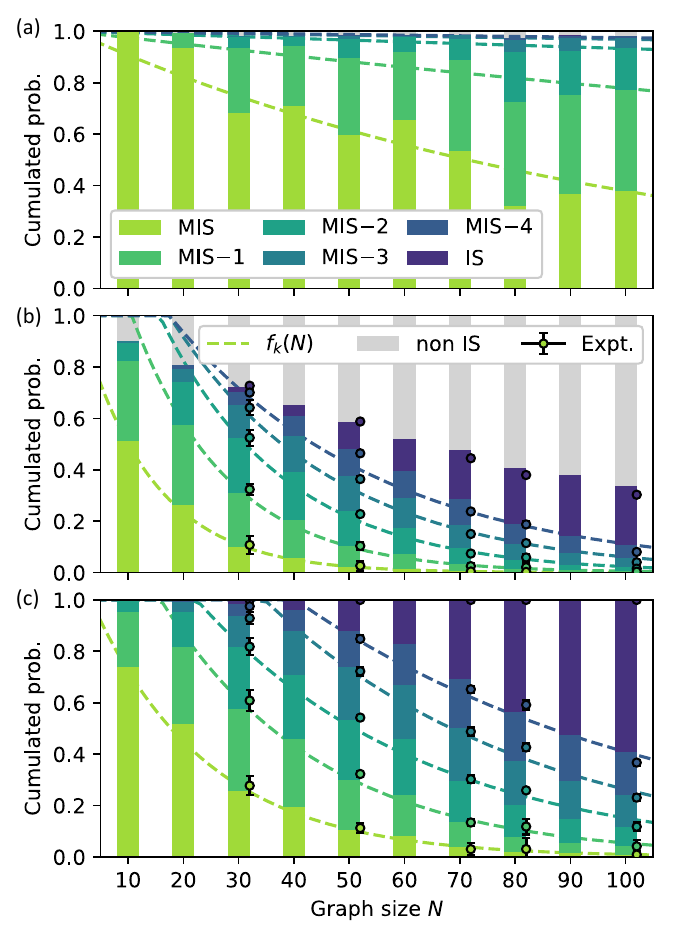}
    \caption{\textbf{Scaling of the preparation efficiency of  the transferable annealing protocol with the graph size.} (a) Cumulative probabilities of finding MIS (light green), MIS$-k$ for $k=1,\dots,4$, IS (dark blue), and non-IS configurations (light gray) obtained by sampling distributions from classical MPS emulation of noiseless annealing. Probabilities are averaged over $10$ graphs of the same size. The decline of $\sum_{i \leq k}$P(MIS-i) with increasing graph size $N$ is fitted with exponential decay (dashed lines in corresponding colours), with decay constants $N_k^{\mathrm{emu}}$ provided in Table~\ref{tab:decay-constant}. (b) Same as (a), but with imperfect sampling ($\varepsilon=3\%$ and $\varepsilon^\prime=8\%$) applied to benchmark experimental noise as described in App.\,\ref{app:spam}. The results are compared with experimental data (dots) obtained on the Orion Alpha device for graph sizes $N=30, 50, 70, 80,$ and $100$. The decay is fitted similarly to (a), with decay constants $N_k^{\mathrm{expt}}$. (c) Same as (b), but after applying the post-processing procedure described in App.~\ref{app:post-process}. The decay is fitted similarly to (a), with decay constants $N_k^{\mathrm{proc}}$.
 }
    \label{fig:proba-scaling}

\end{figure}

\subsection{Expected scaling of MIS-k probabilities}
\label{ssec:scaling-num}
To further evaluate the quality of the states prepared at larger system sizes, we analyse the probability distributions of configurations for each graph. Fig.\,\ref{fig:proba-scaling}(a) shows the average cumulative probabilities for various types of configurations: non-independent sets (non-IS), independent sets (IS, with the largest labelled as MIS$-k$, indicating they are missing $k$ elements to qualify as MIS), and MIS. While for unseen graphs of the same size as those used for training, i.e. $N=10$, the MIS probability remains close to $1$, at $N=100$, the average MIS probability lies around $37\%$. Notably, the probability of finding a MIS decreases with increasing graph size, as theoretically expected. 

Indeed, the performance of the quantum annealing procedure can be effectively described by quasi-adiabatic evolution, where the Landau-Zener transition probability out of the ground state decreases exponentially with the minimum energy gap encountered during the process. At best, the system undergoes an Ising-type second-order quantum phase transition~\cite{sachdev1999quantum} from a disordered initial state to the ordered MIS phase~\cite{bernien2017probing}, with a polynomial gap scaling as $\Delta \approx N^{-\alpha}$,
where $\alpha = 1/2$~\cite{schuler2016universal}. However, this argument holds strictly for instances where the MIS is non-degenerate and deriving a careful scaling expression for P(MIS-k) is out of the scope of this paper.  
Even in cases where MISs were not found, a significant fraction of the probability distribution remains concentrated in MIS$-1$ and MIS$-2$ configurations. To understand the decay of the summed probabilities over MIS$-k$ configurations with graph size, we can nonetheless perform a naive fit using a piecewise exponential decay model, 
\begin{equation}
    f_k(N) =
    \left\{
    \begin{array}{ll}
        \quad\quad\quad 1 & \text{if } N \leq b_k, \\
        \exp\left(-\frac{N-b_k}{N_k}\right), & \text{else}.
    \end{array}
    \right.
\end{equation}
The decay constants extracted from the emulated data are summarised in Table\,\ref{tab:decay-constant}. Following these fits, we can extrapolate that measuring a MIS with $F=99\%$ probability at size $N=500$ would require around $n_{\rm shots}\geq\log(1-F)/\log(1-\exp(-N/N_0^{\rm emu}))\approx 600$ shots. However, relaxing the quality and measuring a MIS or a MIS$-1$ would only need around $14$ shots with this method. This feature is particularly promising for integrating the VQAA approach into more sophisticated hybrid algorithms~\cite{PhysRevA.110.012434,dalyac2024graph}, where suboptimal solutions can be used as warm starts for the classical solver. Therefore, in this noiseless numerical scenario, the transferable protocol proves useful for optimisation pipelines addressing large instances where obtaining sufficiently good enough solutions from the QPU remains valuable. 

\input{parameters_table}

\subsection{Experimental scaling and classical post-processing}
\label{ssec:scaling-exp}
The previous study was performed in the noiseless case, where only diabatic errors due to gap vanishing with the system size can reduce the performance. A standard approach to counterbalance this issue is to extend the schedule duration accordingly, following the adiabatic condition given in App.\,\ref{app:adiabatic} and as demonstrated in Fig.\,\ref{fig:noise-analysis}(b). However, on current noisy hardware with limited coherence time $T_2\approx4.5~\mu$s, elongated dynamics can exacerbate the decoherence impact, necessitating a balance to identify an optimal duration. We therefore stick to the unchanged control schedule for the experimental implementation. We experimentally run the protocol on a reduced part of the test set (graphs with $N=30,50,70,80, 100$) to assess how hardware errors might affect the previous scaling. Fig.\,\ref{fig:proba-scaling}(b) displays the experimental cumulated probabilities obtained for various sizes, again sampling each graph around $n_{\rm shots}=1000$ times. The right panel of Fig.\,\ref{fig:fig_1}(b) depicts a successful MIS shot measured on the device. Overall, the correct MIS sizes can still be retrieved, with deviations up to one or two nodes for the largest graphs. However, the quality of the distributions has significantly declined, with P(MIS) dropping to just $0.05\%$ at $N = 100$. 

This experimental data can be benchmarked, with reasonably good agreement, with previous numerical results, this time imperfectly sampled using the same values of $\varepsilon$ and $\varepsilon^\prime$ as in Sec.\,\ref{sec:applications}. Hence, at these sizes, detection errors dominate: for instance, a perfect preparation of a non-degenerate MIS would be scaled down by a factor $(1-\varepsilon)^{N-S_\Graph}(1-\varepsilon^\prime)^{S_\Graph}\approx0,5\%$ for the given values. As in Sec.\,\ref{ssec:scaling-num}, we fit the seemingly exponential decay of the experimental P(MIS) with the associated decay constants summarised in Table~\ref{tab:decay-constant}. The P(MIS) extrapolated for $N=500$ is now close to $0$ and the quality of solutions that one can hope to retrieve after few hours of sampling has downgraded to MIS-6. Reducing $\varepsilon$ and $\varepsilon^\prime$ remains critically important, though it merely shifts the issue to larger system sizes, such as $N=1000$. 

Thus, MIS-solving methods, regardless of their expected numerical efficiency, must be complemented by classical post-processing steps when implemented on current noisy hardware. These steps can be utilised to reconstruct an optimal solution as described in App.\,\ref{app:post-process}. Specifically, by removing constraint-violating nodes in non-IS and randomly adding new nodes to the identified IS, near-optimal solutions can be classically refined, as verifying whether a modified IS of size $(S_\Graph-1)+1$ ($(S_\Graph-2)+2$) qualify as MIS would only scale with $N$ ($N^2$). Fig.\,\ref{fig:proba-scaling}(c) displays the application of this method to the distributions (emulated with detection errors and raw experimental ones) given in Fig.\,\ref{fig:proba-scaling}(b). The more nodes the post-processing step is allowed to add to an IS, the better the improvement in the measured distribution, as showcased in Fig.\,\ref{fig:postprocess-MIS}. Removing on average $5$ nodes in the non-IS configurations and improving the IS by at most $2$ nodes, we are able to experimentally find the MIS at least once for each graph, even improving on the noiseless emulated case of Sec.\,\ref{ssec:scaling-num}. The cumulative probabilities of the post-processed experimental data decay approximately half as fast as those of the raw data (see Table~\ref{tab:decay-constant}), underscoring the enhancement in the distributions. However, this straightforward approach incurs a classical computation cost. For instance, at $N=100$, removing nodes from non-IS takes approximately $600~$ms, while adding one node to expand an IS requires around $300~$ms, and adding a second node further increases the time by an additional $1.8~$s. When comparing a hybrid quantum-classical approach like this one to a state-of-the-art classical method, it is essential to account for the total time to solution, including both quantum sampling and classical post-processing durations. By utilising the pre-trained protocol transferable to larger triangular graphs, we can successfully sample MIS configurations up to $N=100$ within few tens of seconds, orders of magnitude below what a closed loop run on hardware would require but sill three orders of magnitude below the state-of-the-art CPLEX method. 

\section*{Conclusion and outlook}
We introduced a novel transferability method for quantum annealing that demonstrates promising performance on systems with up to $100$ qubits. By training the protocol on a dataset of small graphs with similar structures, we showed that the same control schedule could be effectively applied to larger instances with analogous features. The optimisation process accounts for realistic hardware constraints and avoids regions with vanishing energy gaps, ensuring practical applicability.
 
Our approach was tested on real-world smart-charging problem instances derived from EDF data, yielding encouraging results. The method leverages the fact that all graphs in the dataset are subgraphs of a shared atomic layout. However, this dependence poses a limitation for encoding larger industrial instances, which often involve significantly more constraints than the atomic layout can accommodate. In this work, we focus exclusively on graphs with relatively low connectivity, as natively embedding graphs with higher connectivity or non-local edges remains challenging. A potential approach for handling such non-UD graphs involves the use of gadgets that require a quadratic number of auxiliary qubits~\cite{Nguyen22}, but this approach is prohibitively costly. A more practical alternative would be to utilize similar gadgets while transforming only the minimal non-UD structures of the graph. By relaxing the validity conditions for specific weighted graph structures, it becomes possible to reduce the number of required qubits. This reduction enables the derivation of an efficient embedding on a triangular graph~\cite{DelpechBM24}. Once gadgetised, these graphs can typically mapped onto regular lattices of traps, and the efficiency of applying a transferable protocol specific to this regular lattice could be studied in the future. 
To evaluate scalability, we generated a dataset of graphs up to $N=100$. While the probability of obtaining low-energy bitstrings decreases exponentially with system size, our fitted model reveals that the decay rate for near-optimal solutions scales favourably. This suggests that quantum annealing could serve as an effective low-energy sampler, especially for approximate solutions to complex problems. Combined with recent evidence of scaling advantages over classical methods~\cite{bauza2024scaling} these results hint at the growing potential of quantum hardware to provide a meaningful advantage in generating high-quality approximate solutions for industrially relevant problems with limited resources.

\section*{Acknowledgments}
The authors thank Pasqal hardware department, including Guillaume Villaret, and the software platform team for their continuous work, making this project achievable. This work was developed as a collaboration initiated within the European Commission Quantum Flagship projects PASQuanS (817482) and PasQuans2 (10113690). This work was made possible thanks to the Pack Quantique grant from région Ile de France and GENCI, project AQUARE (convention N°20012757). 

\bibliographystyle{unsrtnat}
\bibliography{biblio}

\onecolumngrid
\appendix
\renewcommand{\thefigure}{A\arabic{figure}}
\setcounter{figure}{0} 
\newpage
\input{appendices}

\end{document}

%% file: parameters_table.tex
\begin{table}
\centering
\begin{tabular}{c|c|cc|cc|}
\cline{2-6}
                            & \textbf{Emulated}  & \multicolumn{2}{c|}{\textbf{Experimental}}                 & \multicolumn{2}{c|}{\textbf{\begin{tabular}[c]{@{}c@{}}Expt. \\ post-processed\end{tabular}}} \\ \cline{2-6} 
                            & $N_k^{\rm emu}$    & \multicolumn{1}{c|}{$N_k^{\rm expt}$}   & $b_k^{\rm expt}$ & \multicolumn{1}{c|}{$N_k^{\rm proc}$}                     & $b_k^{\rm proc}$                  \\ \hline
\multicolumn{1}{|c|}{$k=0$} & $1.0(1)\times10^2$ & \multicolumn{1}{c|}{$1.3(0)\times10^1$} & $1$              & \multicolumn{1}{c|}{$2.1(1)\times10^1$}                   & $3$                               \\ \hline
\multicolumn{1}{|c|}{$1$}   & $4.0(4)\times10^2$ & \multicolumn{1}{c|}{$1.7(0)\times10^1$} & $11$             & \multicolumn{1}{c|}{$2.9(1)\times10^1$}                   & $16$                              \\ \hline
\multicolumn{1}{|c|}{$2$}   & $1.4(1)\times10^3$ & \multicolumn{1}{c|}{$2.2(1)\times10^1$} & $16$             & \multicolumn{1}{c|}{$4.1(2)\times10^1$}                   & $22$                              \\ \hline
\multicolumn{1}{|c|}{$3$}   & $3.2(3)\times10^3$ & \multicolumn{1}{c|}{$2.9(2)\times10^1$} & $17$             & \multicolumn{1}{c|}{$4.8(6)\times10^1$}                   & $35$                              \\ \hline
\multicolumn{1}{|c|}{$4$}   & $4.0(5)\times10^3$ & \multicolumn{1}{c|}{$3.8(3)\times10^1$} & $17$             & \multicolumn{1}{c|}{$6.7(7)\times10^1$}                   & $40$                              \\ \hline
\end{tabular}
\caption{\textbf{Decay parameters used to fit emulated , raw experimental and post-processed experimental cumulative probabilities, shown in Fig.\,\ref{fig:proba-scaling}.}}
\label{tab:decay-constant}
\end{table}

%% file: appendices.tex
\section{Mapping MIS problems to neutral atoms}
\label{app:mapping}

\begin{figure}[b]
    \centering
    \includegraphics[width=\textwidth]{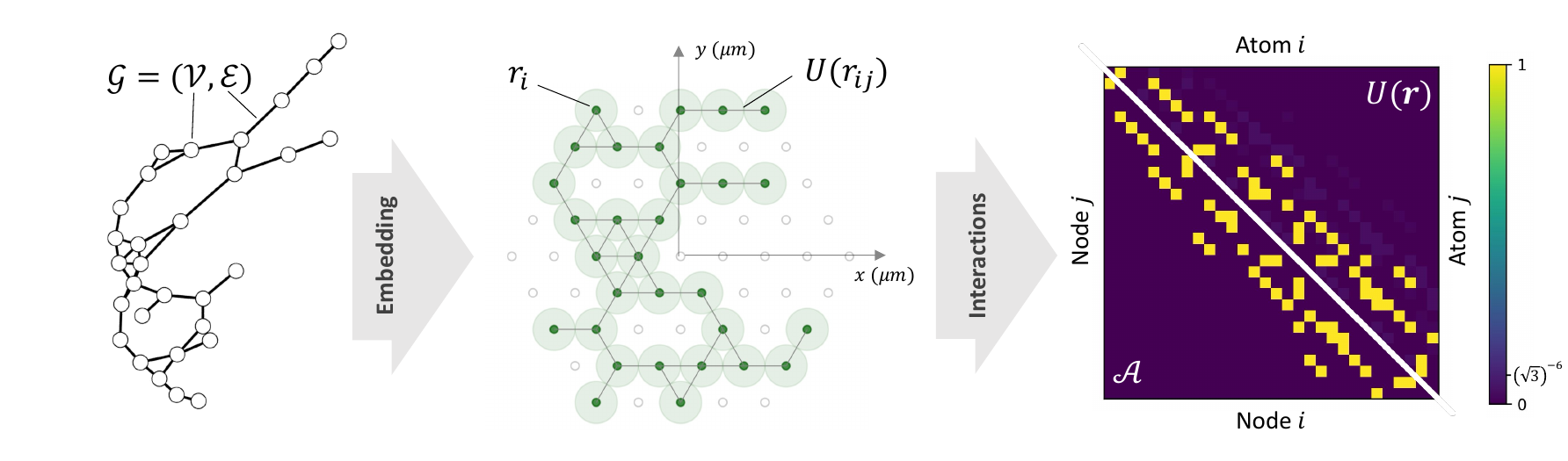}
    \caption{\textbf{Representing graph connectivity with interactions.} With an adequate spatial positioning $\boldsymbol{r}$ of the atoms in a register (green), their interactions (upper corner of the matrix) can reproduce the adjacency matrix $\mathcal{A}$ (lower corner of the matrix) of an input graph $\Graph$, up to a global scaling factor. The sharp decay of the interaction allows neglecting terms between second (and further) nearest neighbours.}
    \label{fig:embedding-graph}
\end{figure}

The topology of the interaction Hamiltonians obtained with Rydberg atoms natively echoes the ones of the classical spins models, making the neutral atom technology a potential resource to tackle specific graph problems. 
By utilising the spatial dependency of the two body terms in $\hat U(\boldsymbol{r})$, one can for instance reproduce the adjacency matrix $\mathcal{A}$ of a graph $\Graph$ as shown in Fig.\,\ref{fig:embedding-graph}. We define the corresponding graph Hamiltonian as, 
\begin{equation}
\hat{H}_\Graph=\sum_{(i,j)}\mathcal{A}_{ij}\hat{n}_i\hat{n}_j.
\end{equation}
More precisely, assigning qubits to vertices enables a direct correspondence between their pairwise interactions and the weights of the edges. Mapping a $N$-node graph to an interaction Hamiltonian thus requires placing $N$ qubits such as to satisfy the $N(N-1)/2$ constraints $U(r_{ij})=\mathcal{A}_{ij}$ $\forall (i<j)$ up to a global scaling coefficient. The task of solving this \textit{embedding} problem poses challenges in itself, and the subsequent sections will delve into various options for resolution.

As a reminder, the Rydberg blockade effect can be approximated as enforcing that pairs of atoms can not be both excited at the same time, if located closer than a certain threshold $r_b$. Incidentally, distant pairs remain free from this constraint. This binary perspective, solely depending on the threshold $r_b$, natively mimics the topology of UD graphs where two vertices $i,j$ are sharing an edge if and only if they lie within a threshold distance from each other in the Euclidean plane.

Thus, embedding UD graphs into a neutral atom system amounts to a free-space positioning problem with $2N$ variables $\boldsymbol{r}=(x_i,y_i)_{i=1\cdots N}$ and the constraints $U(r_{ij})/U(r_b)\gg 1$ for $(i,j)\in\edges$ and $\ll 1$ otherwise. 

The resulting Ising interaction Hamiltonian reads, 
\begin{equation}
    \begin{split}
        \hat U(\boldsymbol{r}) & =\sum_{(i,j)\in\edges}U(r_{ij})\hat{n}_i\hat{n}_j+\sum_{(i,j)\notin\edges}U(r_{ij})\hat{n}_i\hat{n}_j \\
     & = U(r_b)\hat{H}_\Graph+U(r_b)\sum_{(i,j)\in\edges}\left(\frac{r_b^6}{r_{ij}^6}-1\right)\hat{n}_i\hat{n}_j+\sum_{(i,j)\notin\edges}U(r_{ij})\hat{n}_i\hat{n}_j,
    \end{split}
    \label{eq:mapping}
\end{equation} 
where we artificially separate into two the sum over pairs sharing an edge in order to highlight the presence of $\hat{H}_\Graph$. 

The first sum accounts for the variation of distances between pairs representing an edge. For a perfect embedding, all linked pairs are spaced by $r_b$, cancelling the sum. Non-zero terms in that sum can lead to lifts of degeneracy of blockaded states in the spectrum of $\hat U(\boldsymbol{r})$. For instance, a $2\times2$ square with atoms spaced by $a$ represents in a fully blockaded regime, i.e. $r_b>\sqrt{2}a$, the simplest King's graph where all four nodes are connected. While all combinations of highlighting two nodes out of four hold the same cost in the graph, exciting two atoms on the same border or two atoms across a diagonal does not yield the same energy in $\hat U(\boldsymbol{r})$. It can be worth to note when wanting to reproduce perfectly the spectrum, but harmless when only considering ground state mapping. 

The second sum gathers the spurious terms generated by the tail of the interactions, which are usually neglected in the blockade approximation. One has to ensure that the largest distance between a pair sharing an edge in the graph is always far less than the shortest distance between a pair not sharing an edge. For instance, when embedding onto a regular register, one needs to be aware of the ratio between nearest-neighbour (NN) distance $r_{nn}$ and next nearest-neighbour (NNN) $r_{nnn}$ distance. 
For a theoretical proper embedding of a binary adjacency matrix, where all edges have the same weight, the distances should be chosen as $r_{nn}\ll r_b\ll r_{nnn}$, in order for the NN pairs to represent the edges and the interactions between NNN pairs to be neglected.
More practically, a good choice for the blockade approximation to work is $r_b\approx\sqrt{r_{nn}r_{nnn}}$.
However, for regular lattice or arbitrary configurations of atoms, having $r_{nnn}/r_{nn}\gtrapprox 1$ will translate into erroneous blockade approximation and thus smaller energy separation between blockaded and non-blockaded states. Going back to the previous square geometry, this ratio is $\sqrt{2}$. Therefore, embedding a graph corresponding to a square without diagonal connections is especially tricky as the diagonal terms, supposedly not representing edges, will still contribute with strength $1/8$ regardless of the chosen blockade radius. 

Embedding UD graphs into a plane can be achieved in a relative straightforward way using \textit{force-directed graph drawing} methods such as the Fruchterman-Reingold algorithm~\cite{Fruchterman91}. The latter aims at positioning nodes such that connected pairs of nodes are placed close together, while unconnected pairs of nodes are positioned farther apart. The algorithm simulates a physical system where nodes are represented as particles that repel each other due to an electrical charge and are connected by springs. The optimal positions, iteratively obtained, correspond to the equilibrium state reached by the system.

\section{Bayesian optimisation to efficiently navigate a cost landscape}
\label{app:bayesian-optimisation}

Bayesian optimisation algorithm  has two keywords, model and decide \cite{brochu2010tutorial}. It encompasses both a statistical model that reconstructs the landscape of the target function $C$ while providing an uncertainty on such reconstruction and a decision maker, the acquisition function $a$, which indicates where the next evaluation will be most likely to enhance optimisation. After a training phase where the model first fits few evaluations of $C$, each new iteration will favour either exploitation of promising areas of the parameter space or exploration of regions where the high uncertainty leaves room for a potential minimum. The newly acquired measurement $C(\theta)$ updates the prior knowledge of the model using a Bayesian inference technique. 

In order for the prior model to reproduce the target, we need to train it on a few observed evaluations of $C$. Provided with a budget of $n_r$ calls, $C$ is surveyed according to an initial space-filling pattern. For small-sized spaces, the probing points, if enough, can be formed into a discrete grid efficiently covering the entire parameter space. However, as the number of dimensions grows, it becomes less resource consuming to utilise a more efficient space filling pattern such as Latin hypercube sampling. The latter ensures a covering of the entire range of each input variable, while minimising correlation between variables. The acquired training dataset $D=(\thetab,\Cb)=\{\theta^{(k)},C(\theta^{(k)})\}_{k=1,\hdots,n_r}$ will be used to prepare an initial model of $C$.

\begin{figure}[b]
    \centering
    \includegraphics[width=\textwidth]{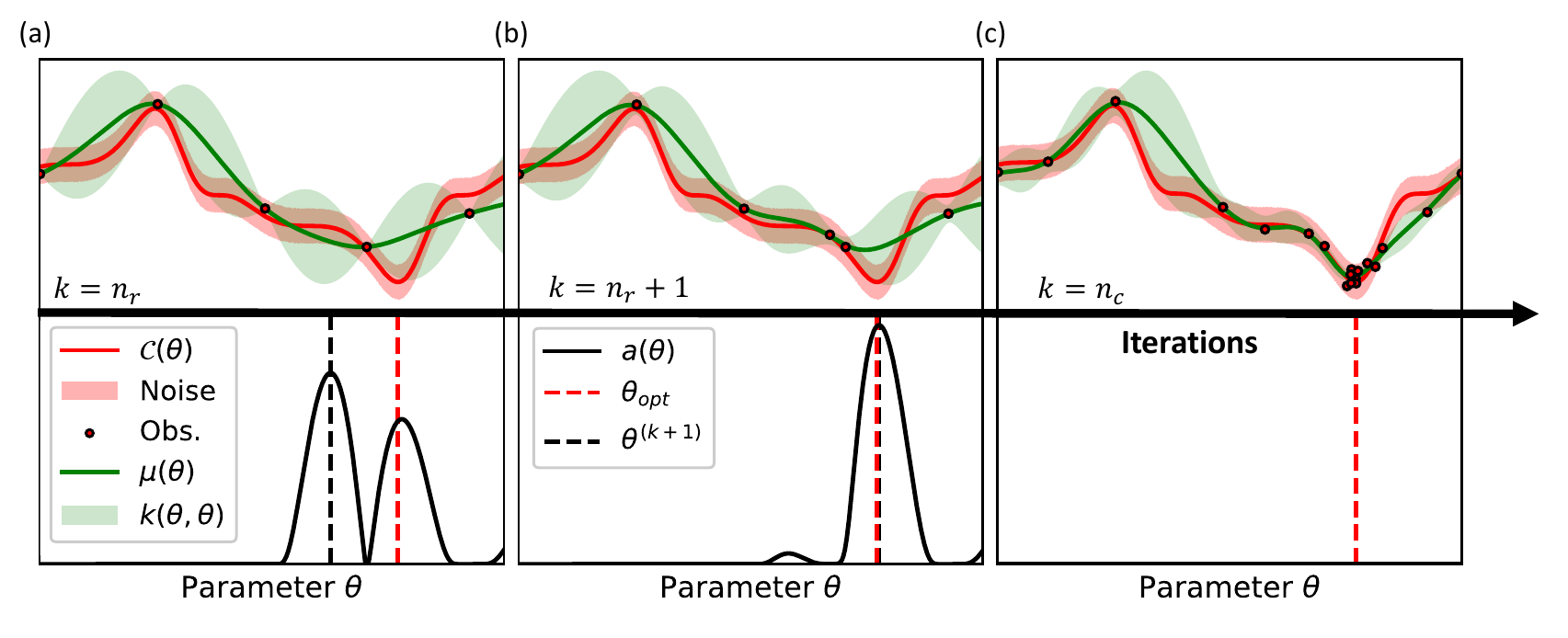}
    \caption{\textbf{Iterative procedure of a Bayesian optimisation.} (a) A noisy cost function $C$ (red) is sampled $5$ times to generate observation data (dots) used for training a Bayesian algorithm. The fitted GP model (green) approximates $C$ with $\mu$ with relative uncertainty given by $k$. Below, the maximisation of the acquisition function $a$ (black) outputs the next point to query $\theta^{(k+1)}$ (dashed black). (b) Updating the model with this new observation modifies $\mu$, $k$ and $a$ iteratively. (c) After enough iterations, the model has reproduced interesting regions of $C$ and locates the minimum $\theta_{\rm opt}$ (dashed red). }
    \label{fig:bo-ex}
\end{figure}

An example is given in Fig.\,\ref{fig:bo-ex}(a) where a noisy $C(\theta)$ (red) is approximated by a Gaussian process, using a fitted Matern kernel, trained over $5$ observations spaced with the latin hypercube sampling method. Regions devoid of observations exhibit large uncertainties and since the points are somehow distant, the noisy behaviour of the function has yet to be understood by the model. 

The use of the Bayesian algorithm seems justified to study unknown functions with many local minima, as long as the cost evaluation remains large in front of the model updating duration (few seconds after $100$ iterations). The algorithm is thus suited as a classical routine for VQAs. For cases where the cost evaluation is negligible, other gradient-free methods are preferable. 

\section{Adiabatic theorem}
\label{app:adiabatic}
Let $\ket{z_0(s)}, \ket{z_1(s)}$ be the instantaneous ground-state and first excited state of $H(s)$, with respective energies $E_0(s)$ and $E_1(s)$. According to the adiabatic condition\,\cite{amin2009consistency}, a quantum system remains in its instantaneous ground-state if the following condition on the total time $T$ is satisfied:
\begin{equation}
\label{eq:adiabatic_cond}
    T \gg \frac{1}{\varepsilon} \max_{s \in [0, 1]} \frac{|\langle z_0(s)|\frac{d}{ds}H(s)| z_1(s) \rangle |}{|E_0(s) - E_1(s)|^2}.
\end{equation}

This condition ensures that the probability of not finding the system in the ground-state is at most $\varepsilon^2$. More formal versions of the adiabatic conditions exist, and we refer the interested reader to the review by Albash and Lidar\,\cite{albash2018adiabatic}, but this version is sufficient to put forward the important condition for adiabaticity. Note that the term $|\langle z_0(s)|\frac{d}{ds}H(s)| z_1(s) \rangle |$ can be seen as the difference in slopes between the ground-state and the first excited state. Because it has no singular scaling with the system size $n$\,\cite{bapst2013quantum}, the adiabatic condition of equation\,(\ref{eq:adiabatic_cond}) can be replaced by the simpler form 
\begin{equation}
\label{eq:gap}
    T \gg \mathcal{O}(N \Delta^{-2}),
\end{equation}
where $\Delta = \min_{s \in [0,1]}|E_0(s) - E_1(s)|$. This means that the total time of the adiabatic protocol is governed by the minimum gap $\Delta$ and its scaling with $n$. Typically, we do not expect to find $\Delta=0$ for non-degenerate Hamiltonians which would correspond to a level crossing between the ground-state and the first excited state during an adiabatic evolution. The intuition behind this can be built by looking at the adiabatic process on a simple two-level generic Hamiltonian that depends on $s$
\begin{equation}
    H(s) = \begin{pmatrix}
        a(s) & c(s) + id(s) \\
        c(s) - id(s) & b(s) \\
    \end{pmatrix},
\end{equation}
where $a,b,c$ and $d$ are real-valued functions. In this case, the level crossing corresponds to having $a(s) = b(s)= x$ and simultaneously $c(s)=d(s)=0$.  The parameterised curve $(a(s),b(s),c(s),d(s))$ in $\mathbb{R}^{4}$ does not possess any specific characteristics or constraints that would cause it to cross $(x, x, 0, 0)$. On the other hand, if $H(s)$ commutes with $\hat{\sigma}_x$, then $a(s)=b(s)$ and $d(s)=0$, and it seems more probable that for some $s$, $c(s)=0$ which would correspond to level-crossing. These arguments can be generalised to Hamiltonians acting on $n$ qubits and the authors of\,\cite{Farhi00} argue that in the absence of symmetries, levels typically do not cross during an adiabatic evolution. Although levels do not cross, the \textit{anti-crossing} phenomenon can create exponentially small gaps that make the evolution time impractically large.

\section{Avoiding diabatic errors with an optimised annealing drive}
\label{app:vQAA}
\begin{figure*}[!]
    \centering
    \includegraphics[width=\textwidth]{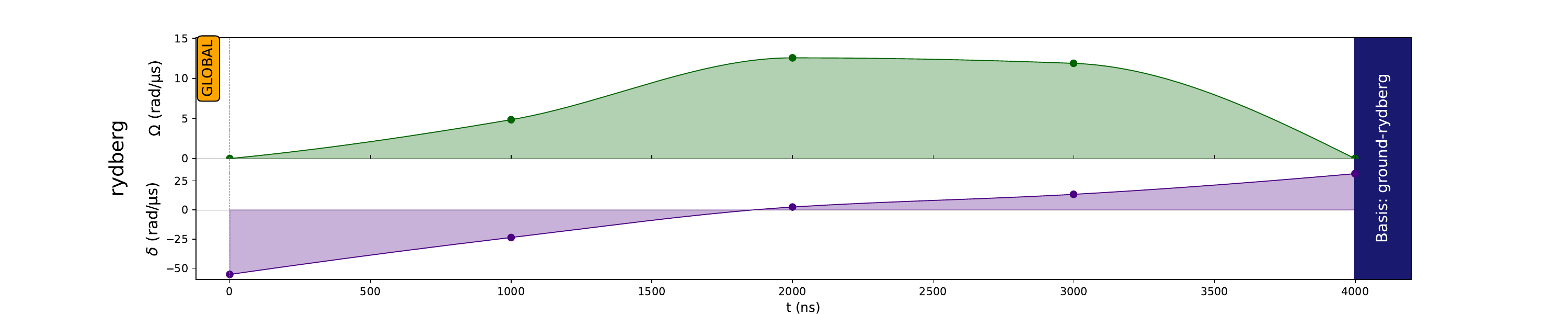}
    \caption{\textbf{Example of an annealing protocol used on an analog neutral atom quantum processor}. Both global control fields $\Omega(t)$ (lightgreen) and $\delta(t)$ (purple) are parameterised by their values at fixed points in time and interpolated in-between.}
    \label{fig:optim-pulse}
\end{figure*}

\begin{figure*}[!]
    \centering
    \includegraphics[width=\textwidth]{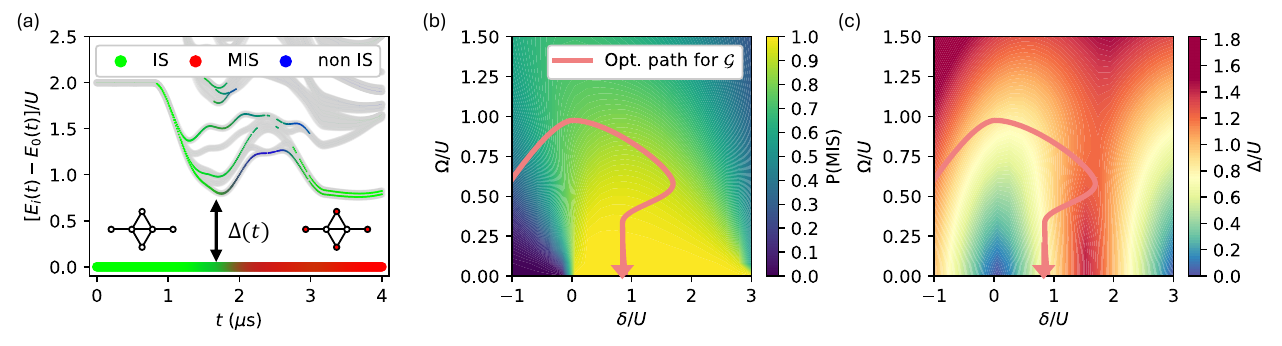}
    \caption{\textbf{Optimisation results when applying a VQAA protocol optimised on a $6-$qubit graph} (a) Instantaneous energy spectrum of the $6-$qubit system displayed during an optimised annealing dynamics, with energy levels colour-coded by their proportions of non-IS (red), IS (green), and MIS (blue) configurations, and line thickness representing population in that level. Ground and first excited states are separated by a gap $\Delta(t)$. (b) Phase diagram of the MIS probability obtained by computing the ground state of the Rydberg Hamiltonian for the previous graph with constant parameters $\Omega/U$ and $\delta/U$. The path taken by the optimised protocol (red) is constrained by $\Omega/U\leq1.5$ and $|\delta|/U\leq3$. (c) Similar as (a) for the gap $\Delta$. The optimised protocol avoids regions of vanishing gaps (blue).}
    \label{fig:optim-unique}
\end{figure*}

To demonstrate the strength of the VQAA described in Sec.\,\ref{sec:ingredients}, we seek to find an optimised path quasi-adiabatically driving a system towards the ground state of a $6-$qubit system, reproducing the graph showcased in Fig.\,\ref{fig:optim-unique}(a). We diagonalise the Rydberg Hamiltonian over $\Omega/U\in[0,1.5]$ and $\delta/U\in[-1,3]$ and plot in Fig.\,\ref{fig:optim-unique}(b) the MIS probability obtained when sampling the ground state. A lobe-shape portion of the phase diagram (yellow) constrained between $\delta/U=0$ and $\delta/U=3$ exhibits MIS probabilities higher than $95\%$. We emulate an optimisation procedure with smooth pulse shaping using a periodic temporal slicing with $m=4$ over the parameter space $\Theta=[0.5\,\mu$s$, T]\times[0,1.5\,U]^m\times[-2\,U,3\,U]^{m+2}$ and use a total of $n_{\rm iter}=10+90$ iterations with perfect access to the states produced. We set a maximum duration of $T=3~\mu$s for the protocols explored by the optimiser, compelling it to manipulate the shapes of $\Omega$ and $\delta$ to ensure adiabaticity within these constraints. The gap $\Delta$ is defined between the MIS manifold of states (here only including the ground state, since the graph has only one MIS) and the first above excited state. 

With the optimised path displayed in Fig.\,\ref{fig:optim-unique}(b), we obtain a MIS probability close to $1$ after $3\mu$s. While the optimiser has hit the temporal bound, hinting at the fact that longer protocols could be even more successful, both amplitude and detuning were used with parsimony, without any explosion of the parameters. Although the path may initially appear convoluted, Fig.\,\ref{fig:optim-unique}(c) illustrates that it effectively circumvents regions characterised by vanishing energy gaps, i.e. $\Omega/U\approx0$ and $\delta/U\approx 0/3$, thus trying to maintain adiabaticity, as excessively small values might induce undesired population transfers to excited states. Indeed, the evolution duration $T$ dictates the minimum acceptable gap value (see App.\ref{app:adiabatic}), here around $0.7U$. The protocol is in fact not perfectly adiabatic, as shown by the instantaneous energy spectrum of Fig.\,\ref{fig:optim-unique}(a). While most of the population remains in the ground state, transitioning from the IS phase to the MIS phase, a bit of population leaks to the first excited state around $1~\mu$s  when the gap is minimum. The Bayesian optimizer typically avoids regions susceptible to diabatic errors. At the same time, it can discover paths where early unwanted population transfers are later recovered, returning to lower-energy states by the end of the evolution, effectively finding shortcuts to adiabaticity. This constitutes a strength of the method, as it reduces the need to remain perfectly adiabatic throughout the whole dynamics, which would require elongated protocols.

\section{QAOA framework}
\label{app:qaoa}
\subsection{Formalism}

The Quantum Approximate Optimisation Algorithm (QAOA)~\cite{Blekos_2024} is a prominent method within the quantum optimisation paradigm~\cite{QAOA,Zhou20}, inspired by the Trotterised version of adiabatic evolution. In QAOA, the quantum unitary evolution operator is parameterised by two sets of $p$ angles, $p$ being the \textit{depth} of the algorithm, and alternates between two types of operations: mixing and problem-specific evolution. \\

    \textbf{Hardware-agnostic version:} The algorithm applies $p$ successive layers of two distinct Hamiltonians: a cost operator $\hat{C}$ and a mixing operator $\hat{M}$. Mathematically, this can be represented as:
    \begin{equation}
        \ket{\psi(\thetab)} = \prod_{j=1}^p \hat{U}_M(\thetab_{2j}) \hat{U}_C(\thetab_{2j+1}) \ket{\psi_0}.
    \end{equation}
    $\hat{U}_{X}(\theta) = e^{-i\theta \hat{X}}$ for $X\in\{M,C\}$ denotes the unitary operators corresponding to mixing and cost evolution, respectively. The algorithm initialises by preparing the system in $\ket{\psi_0} = \ket{+}^{\otimes N}=2^{-N/2} \sum_{\boldsymbol{z}\in\mathbb{B}^N} \ket{\boldsymbol{z}}$, representing the ground state of the mixing Hamiltonian, often chosen as $\hat{M} = \sum_i \hat{\sigma}_i^x$. By evolving from this state, and with the aid of a classical optimiser, it becomes possible, as $p \rightarrow \infty$, to determine optimal angles $\thetab^*$ such that:
    \begin{equation}
        |\psi(\thetab^*)\rangle = \text{argmin}_{\psi}\bra{\psi}\hat{C}\ket{\psi}
    \end{equation}
    This approach can be viewed as a quantum version of annealing using quantum tunnelling to locate the minimum of a landscape.\\
    
    \textbf{Rydberg version:} In the Rydberg version, the cost operator $\hat C$ can be replicated for specific graphs and the mixing part can be tackled with resonant pulses of amplitude $\Omega$. A major limitation preventing a straightforward implementation of QAOA on a Rydberg setup lies in the inability to turn off the interaction component of the Rydberg Hamiltonian during the mixing phase of each layer. This also implies that a proper preparation of $\ket{\psi_0}$ might require a more complex protocol or be even not possible. Applying pulses with $\Omega\gg U$ enables to neglect the interaction effects, but the maximum amplitude reachable on current devices is only around $\Omega_{\rm max}/2\pi\sim$ few MHz. Consequently, this requirement mandates low values of $U$ and hence of $\delta$, which in turn prolongs the sequence and renders the evolution vulnerable to decoherence. One can nonetheless program an evolution with a QAOA-like protocol, applying series of resonant pulses with fixed $\Omega$ interleaved by periods of free evolution under fixed detuning $\delta$. Each layer duration constitutes a pair of parameters to optimise on such that $\thetab=\left\{t_{\rm cost}^j,t_{\rm mix}^j\right\}_{j=1,\hdots, p}$. 

\begin{figure}[h]
    \centering
    \includegraphics[width=\textwidth]{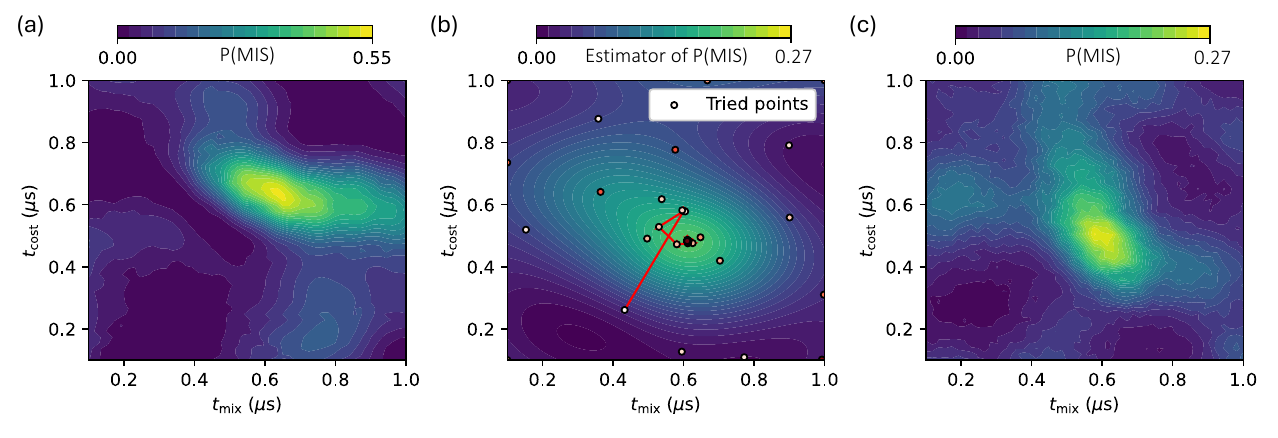}
    \caption{{\textbf{Landscape of MIS probability for QAOA with depth $1$.} (a) Varying the two parameters $t_{\rm mix}$ and $t_{\rm cost}$ of a QAOA-like sequence with $p=1$, the noiseless landscape browsed when optimising P(MIS) can be retrieved. (b) The Bayesian optimiser constructs an estimator of P(MIS) by strategically sampling the parameter space (dots) with the help of the QPU. The redder the dot, the later it has been sampled. The minimum is updated (red line) each time a better one is found. (c) Adding detection and calibration errors in the emulation moves the global minimum location in the parameter space and explains the model built by the optimiser when using the noisy QPU.}}
    \label{fig:QAOA-landscape}
\end{figure}
\subsection{Closed loop and miscalibration}
\label{app:miscalibration}

We apply the QAOA-like approach as a way to benchmark both the ability of the QPU to perform quantum dynamics under such  protocols and the ability for the Bayesian optimiser to find optimal driving parameters even in the presence of noise. 

We perform a closed loop with $p=1$ on the $N=6$ graph of Fig.\,\ref{fig:optim-unique}(a), using the MIS probability as a figure of merit in this case. The landscape of optimisation in the noiseless case is displayed in Fig.\,\ref{fig:QAOA-landscape}(a) with a unique global optimum located at $t_{\rm mix}^*=t_{\rm cost}^*=0.633~\mu$s. We explore this landscape with the Fresnel QPU using a Bayesian optimiser provided with $10+50$ iterations with $256$ shots each. The optimiser starts by using its initialisation budget to randomly sample the parameter space in an efficient covering, as shown by clearer points in Fig.\,\ref{fig:QAOA-landscape}(b). Building a model to estimate the MIS probability at unexplored points, it then exploits the central region (green), sometimes exploring a bit outside of it (red points) to reduce its uncertainty and ultimately converging towards a believed optimum $t_{\rm mix}^*=0.611~\mu$s and $t_{\rm cost}^*=0.483~\mu$s (darker red points). 

Surprisingly, the found optimum and constructed estimator  do not coincide with the noiseless case. We checked that detection errors alone do not move the optimum, revealing that a miscalibration of parameters is involved. Indeed, when measuring $\Omega$ and $\delta$ with independent Rabi and Ramsey protocols, we obtain $\Omega/2\pi=1.08~$MHz and $\delta/2\pi=-0.775~$MHz, instead of $1~$MHz and $-0.5~$MHz, respectively. Fig.\,\ref{fig:QAOA-landscape}(c) exhibits the noisy landscape computed with detection errors and those miscalibrations, highlighting a displacement of the optimum location and a reduced optimal value of MIS probability in accordance with the results found on the QPU. In the presence of limited noise, the Bayesian optimiser can still navigate the modified landscape and helps construct an optimised protocol to sample MIS configurations. This is promising as noise levels on QPU are, although not negligible, often maintained below known thresholds. However, a downside of such an optimised protocol is that it becomes useless once the QPU is recalibrated. Another optimisation procedure needs therefore to be computed, increasing the total cost. Consequently, this stresses the significance of building a VQA able to generate protocols with greater noise resilience and \textit{transferability}.

\section{Robustness of VQAA}
\label{app:noise-VQAA}

\begin{figure}[h]
    \centering
    \includegraphics[width=\textwidth]{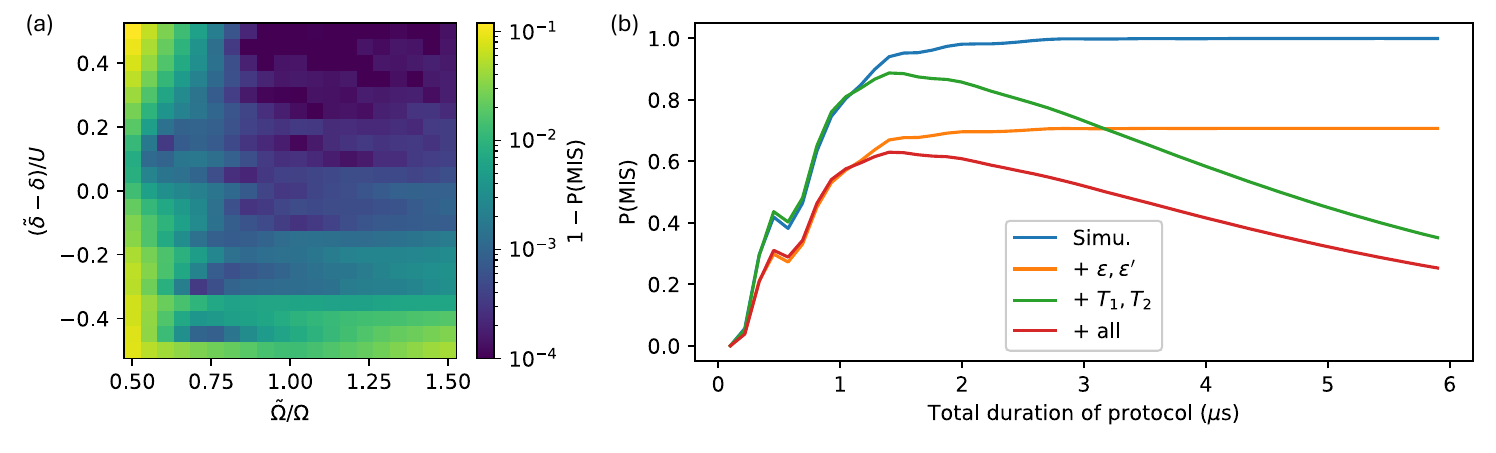}
    \caption{\textbf{Noise analysis of an adiabatic protocol.} (a) Colour map of $1-$P(MIS) when rescaling the global amplitude drive $\Omega$ to $\tilde \Omega$ and shifting the detuning $\delta$ to $\tilde \delta$. (b) Emulated evolution of the MIS probability when stretching an adiabatic protocol in time. In the noiseless case (blue), only diabatic errors impact the result for short times. Detection errors $\varepsilon=1\%$ and $\varepsilon^\prime=8\%$ (orange) rescales the behaviour, while decoherence $T_1=100~\mu$s and $T_2=4.5~\mu$s (green) impacts the probability for longer sequences. With this noise model (red), an optimal duration can be derived, here around $1.4~\mu$s to balance adiabaticity and decoherence.}
    \label{fig:noise-analysis}
\end{figure}

In order to assess the resilience of such almost adiabatic protocols, we conduct a noise analysis, detailed in Fig.\,\ref{fig:noise-analysis}. The protocol appears resilient to parameter miscalibrations, such as rescaling of $\Omega$ and shifts of $\delta$, as depicted in Fig.\,\ref{fig:noise-analysis}(a). In fact, even with similar miscalibrations observed in the QAOA case ($\tilde{\Omega}/\Omega=1.08$ and $(\tilde{\delta}-\delta)/U=0.03$), the obtained MIS probability would only decrease from $99.98\%$ to $99.95\%$, hinting at minimal deviation from the optimum found. The influence of $T$ is also indicated in Fig.\,\ref{fig:noise-analysis}(b) where noiseless simulation exhibits a significant diabatic drop below $1~\mu$s but rather flat behaviour otherwise. Adding the detection errors only rescales the curve by a factor $\approx 71\%$. However, when considering decoherence with an effective model of relaxation $T_1$ and dephasing $T_2$, lengthening the duration becomes detrimental, and an optimal duration starts to appear for this rather small instance. At $T=3~\mu$s, the MIS probability is already reduced by a factor $2$. For larger graphs, this balanced optimum between adiabaticity and decoherence should move towards longer times, thereby reducing the optimum value of P(MIS) that can be achieved. Despite all this, the VQAA approach remains more resilient than the QAOA-like.

\section{Detection errors during the imaging}
\label{app:spam}
\begin{figure}[h]
    \centering
    \includegraphics[width=\textwidth]{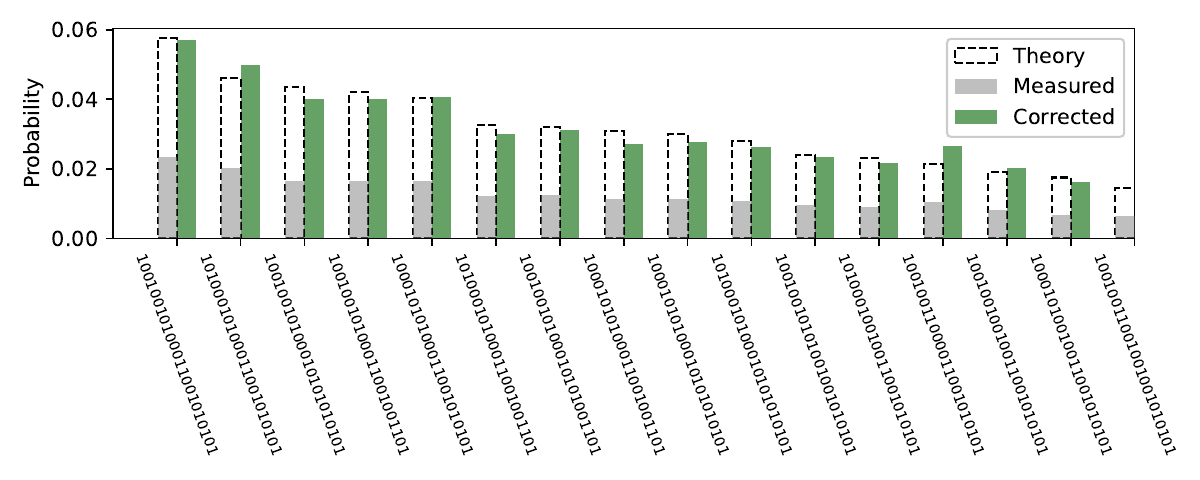}
    \caption{A probability distribution $P_\psi$ (dashed) is incorrectly measured. From the measured distribution $\tilde P_\psi$ (grey) and known values of detection errors ($\varepsilon=3\%$, $\varepsilon^\prime=5\%$), one can approximate the true distribution by a corrected distribution (green) up to $25$ qubits in reasonable time. Only the first $16$ states with the highest probabilities are shown here. }
    \label{fig:spam-histo}
\end{figure}

The measurement process is inherently flawed by several physical processes which can result in measuring an $1$ instead of a $0$, leading to \textit{false positive} detection event and conversely to \textit{false negative}. Background-gas collisions can eject a recaptured atom, emptying a trap and thus flipping a $0$ to an $1$ in a bitstring with probability $\varepsilon\approx 1-3\%$. Additionally, the ejection of atoms in Rydberg states typically lasts around a few microseconds, leaving enough time for some of them to decay from $\ket{1}$, effectively flipping a $1$ to an $0$ in a bitstring with probability $\varepsilon^\prime\approx 5-8\%$. Reducing the temperature of the background atoms help lower $\varepsilon$ while increasing the Rydberg lifetime by using larger Rydberg levels can lower $\varepsilon^\prime$.
The many physical processes inducing bit flips during the measurement phase can be effectively encompassed by two terms, $\varepsilon=p(0\rightarrow 1)$ and $\varepsilon^\prime=p(1\rightarrow 0)$. Keeping in mind that this definition depends on the physical states chosen as $\ket{0}$ and $\ket{1}$, the values of the two terms can vary for various experiments, but are usually at the percent level. Moreover, when the atoms are addressed locally, those values can become site-specific. Modelling these bit flips can be achieved using the following transfer matrix 
\begin{equation}
    M_i=\begin{pmatrix}
        1-\varepsilon_i & \varepsilon_i^\prime \\
        \varepsilon_i & 1-\varepsilon_i^\prime
    \end{pmatrix}
    \label{eq:transfer-matrix}
\end{equation}
As shown in Fig.\,\ref{fig:spam-histo}, the incorrectly measured distribution is thus $\tilde P_\psi=(\bigotimes_i M_i) P_\psi$ where we assume uncorrelated errors. While the detection errors at the single qubit level remain low and easy to correct, they quickly scale with the size of the system. For instance, for $N=100$ and $\varepsilon=1\%$, measuring $\ket{\psi}=\ket{0}^{\otimes N}$ is only achieved with an efficiency of $(1-\varepsilon)^N=36.6\%$. Correcting those errors turns critical for state preparation or algorithmic tasks and requires the inversion of a $2^N\times2^N$ matrix. While the matrix construction/inversion procedure can be sped up using tensor formalism, the most computational resource demanding aspect lies in building the probability distribution vector of size $2^N$. Moreover, due to finite sampling of the state and wrong estimation of $\varepsilon,\varepsilon^\prime$, $(\bigotimes_i M_i^{-1})\tilde P_\psi$ may not be a proper probability distribution. While naive methods such as renormalisation or truncation can give sufficient approximation of $P_\psi$, more advanced methods such as Bayesian reconstruction may prove to yield more accurate results. Such a method is at the moment effectively limited to $N=25$ to remain below a few minutes of computation. 

\section{Classical post-processing of IS distributions}
\label{app:post-process}

A notable feature of NP problems such as the MIS one is that any solution can be verified in polynomial time, meaning that for each measured bitstring, we can quickly verify whether it represents an IS or not. Typically, detection errors introduce noise into the bitstrings, leading to suboptimal solutions. By manually flipping bits, we aim to correct these bitstrings, potentially improving their quality. Specifically, for each bitstring, we check its validity as an IS. If required, we then resolve constraints-violations by ensuring that adjacent nodes are not both included. From a distribution only containing IS, we can iteratively add nodes to form maximal independent sets. The greedy algorithm is parameterised by a depth parameter, which limits the number of nodes that can be added in one recursive exploration. The algorithm ensures that each added node maintains the IS property (i.e., no two adjacent nodes are included) and returns all valid independent sets with the maximum possible size at the given depth. At depth $k$, the time complexity of this greedy algorithm scales as $O(\binom{n}{k})$.

\begin{figure}[h]
    \centering
    \includegraphics[width=\textwidth]{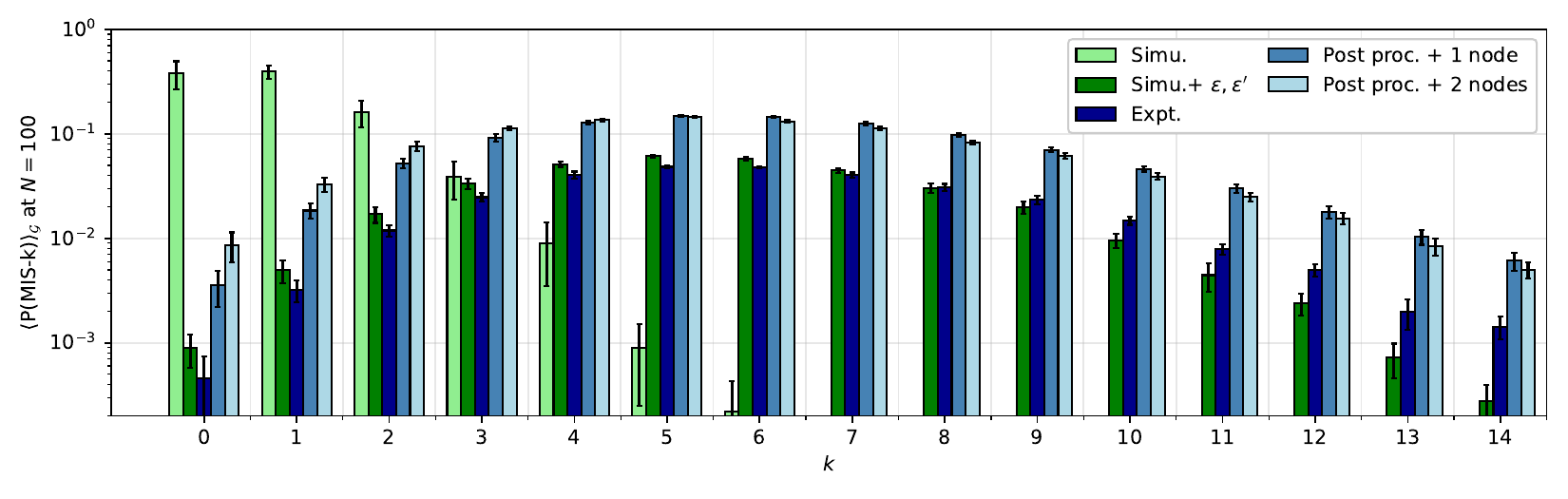}
    \caption{\textbf{Averaged probability distributions of MIS-k instances at $N=100$} are shown for emulated data in the noiseless case (light green) and with detection errors included (dark green), as well as for raw experimental data (dark blue) and post-processed data. The post-processing considers the possibility of adding at most 1 node (blue) or 2 nodes (light blue) to existing IS in order to improve the distribution quality. Additionally, the standard deviation across the 10 graphs of size $N=100$ in the triangular dataset is also plotted.
}
    \label{fig:postprocess-MIS}
\end{figure}

Fig.\,\ref{fig:postprocess-MIS} illustrates the improvement of the measured bitstrings between the raw experimental data and post-processed one with depth parameters $k=1$ and $k=2$. Post-processed data with $k=1$ already yields significantly better results compared to the raw shots, with a nearly ten-fold increase in the probability of obtaining a MIS. This approach was first developed in~\cite{ebadi2022quantum}, where they typically reduced the number of nodes by $3$ to address IS constraints, while adding an average of $1.2$ nodes to maximize the number of selected nodes in the corrected bitstring. In Sec.\,\ref{ssec:scaling-exp}, we reduce by an average of $4.75$ nodes and maximize with $2$ additional nodes. We observe with the case $k=2$ in Fig.\,\ref{fig:postprocess-MIS} that while increasing the depth further still improves the bitstrings, the improvements become more incremental. An interesting direction for future work would be to investigate the trade-off between computational complexity and the depth of the post-processing algorithm.

